\documentclass{IEEEtaes}
\usepackage{amsfonts,bm,amsmath,comment,color,cite,amssymb,subfigure}
\usepackage[]{graphicx}
\usepackage[linesnumbered,ruled]{algorithm2e}
\usepackage{cases}
\usepackage{multicol}
\usepackage{multirow}
\usepackage[section]{placeins}
\usepackage{stfloats}

\def\bzero{{\boldsymbol{0}}}

\def\c1{{\textcircled{a}}}
\def\ba{{\boldsymbol{a}}}
\def\bb{{\boldsymbol{b}}}
\def\bc{{\boldsymbol{c}}}
\def\bd{{\boldsymbol{d}}}

\def\boldf{{\boldsymbol{f}}}
\def\bg{{\boldsymbol{g}}}
\def\bh{{\boldsymbol{h}}}

\def\bq{{\boldsymbol{q}}}
\def\br{{\boldsymbol{r}}}
\def\bs{{\boldsymbol{s}}}

\def\bu{{\boldsymbol{u}}}
\def\bv{{\boldsymbol{v}}}
\def\bw{{\boldsymbol{w}}}
\def\bx{{\boldsymbol{x}}}
\def\by{{\boldsymbol{y}}}
\def\bz{{\boldsymbol{z}}}
\def\bA{{\boldsymbol{A}}}
\def\bB{{\boldsymbol{B}}}

\def\bD{{\boldsymbol{D}}}

\def\bF{{\boldsymbol{F}}}

\def\bI{{\boldsymbol{I}}}
\def\bJ{{\boldsymbol{J}}}

\def\bM{{\boldsymbol{M}}}

\def\bP{{\boldsymbol{P}}}
\def\bQ{{\boldsymbol{Q}}}
\def\bR{{\boldsymbol{R}}}
\def\bS{{\boldsymbol{S}}}
\def\bT{{\boldsymbol{T}}}
\def\bU{{\boldsymbol{U}}}
\def\bV{{\boldsymbol{V}}}

\def\bY{{\boldsymbol{Y}}}

\def\complexC{{\mathbb{C}}}
\def\realR{{\mathbb{R}}}

\def\bzero{{\boldsymbol{0}}}

\def\complexC{{\mathbb{C}}}
\def\realR{{\mathbb{R}}}
\def\expE{{\mathbb{E}}}

\begin{document}
%
\title{Multi-Spectrally Constrained Low-PAPR Waveform Optimization for MIMO Radar Space-Time Adaptive Processing}

\author{Da~Li}
\author{Bo~Tang}
\author{Lei~Xue}
\affil{College of Electronic Engineering, National University of Defense Technology, Hefei 230037, China}


\receiveddate{This work was supported in part by the National Natural Science Foundation of China under Grants  62171450 and 61671453,
and Anhui Provincial Natural Science Foundation under Grant 2108085J30. }

\corresp{{\itshape (Corresponding author: Bo Tang)}.}

\authoraddress{Da Li, Bo Tang, and Lei Xue are with the College of Electronic Engineering, National University of Defense Technology, Hefei 230037, China (e-mail: lida@nudt.edu.cn; tangbo06@gmail.com; eeixuelei@163.com).}

\markboth{Li \MakeLowercase{\textit{et al.}}}%
{Multi-Spectrally Constrained Low-PAPR Waveform Optimization for MIMO Radar Space-Time Adaptive Processing}

\maketitle

\begin{abstract}
This paper focuses on the joint design of transmit waveforms and receive filters for airborne multiple-input-multiple-output (MIMO) radar systems in spectrally crowded environments. The purpose is to maximize the output signal-to-interference-plus-noise-ratio (SINR) in the presence of signal-dependent clutter. To improve the practicability of the radar waveforms, both a multi-spectral constraint and a peak-to-average-power ratio (PAPR) constraint are imposed. A cyclic method is derived to iteratively optimize the transmit waveforms and receive filters. In particular, to tackle the encountered non-convex constrained fractional programming in designing the waveforms (for fixed filters), we resort to the Dinkelbach's transform, minorization-maximization (MM), and leverage the alternating direction method of multipliers (ADMM). We highlight that the proposed algorithm can iterate from an infeasible initial point and the waveforms at convergence not only satisfy the stringent constraints, but also attain superior performance.

\end{abstract}


\section{Introduction}
Multiple-input-multiple-output (MIMO) radar refers to a radar system with multiple transmitters and multiple receivers. Different from traditional phased-array radar, MIMO radar can transmit multiple independent waveforms. Therefore, MIMO radar can leverage the waveform diversity to improve the signal-to-interference-plus-noise-ratio (SINR), operate in more flexible modes, and adapt to the complex environment more intelligently \cite{li2008mimo}. According to the array spacing between the transmitters/receivers, MIMO radar can be categorized into two categories: statistical MIMO radar \cite{Haimovich2008Widely} and coherent MIMO radar \cite{Li2007Colocated}. Statistical MIMO radar has widely separated transmitters/receivers. Therefore, it can fully utilize the spatial diversity to overcome the target fluctuations and improve the target localization accuracy  \cite{fishler2004performance}. Compared with statistical MIMO radar, the transmitters/receivers of coherent MIMO radar are closely spaced. Similar to phased-array radar systems, the transmitters of coherent MIMO radar share the same viewing angle of the targets. Differently, the waveform diversity offered by coherent MIMO radar enables a higher number of degrees of freedom than phased-array radar, resulting in an improved parameter identifiability \cite{Li2007Identifiability}, better target detection performance \cite{ChenVaidyanathan2008MIMOSTAP}, and the capability of supporting multiple functions simultaneously \cite{Tang2022MFRF}.

An airborne early warning (AEW) system (also called AEW and control system), which refers to a radar system operating at a high altitude, is usually used to detect target at a long range. When the AEW system is detecting targets at a low altitude, it might receive strong reflections from, e.g.,  ground. Owing to the AEW platform motion, the ground clutter is extended not only in range and angle, but also in Doppler. Therefore, a weak target is likely to be obscured by mainlobe clutter from the same angle as the target or by sidelobe clutter from different angles but with the same Doppler frequency. These unfavorable factors deteriorate the target detection performance, especially for the slowly moving targets \cite{ward1998space}. To boost the target detection performance in the presence of strong clutter, space time adaptive processing (STAP) techniques have been proposed \cite{ward1998space,Brennan1973stapPaper,Guerci2003stapBook}.  Through collecting waveforms from multiple antennas and multiple pulses, the adaptive multi-dimensional filters of STAP can form deep notches along the clutter ridge and thus suppress the clutter power to a low level.

Considering the superiority of MIMO radar and STAP, researchers proposed the concept of MIMO-STAP for future AEW systems and extensive studies have been devoted to this area (see, e.g., \cite{Mecca2006mimoSTAP,ChenVaidyanathan2008MIMOSTAP,ForsytheBliss2010mimoGMTI,XueLiStoica2010mimoGMTI,cui2021knowledge,wen2021slow} and the references therein).
The results showed that for detection of slowly-moving targets, MIMO-STAP achieved better performance than conventional STAP methods.
However, these studies mainly focused on the design of receivers for MIMO-STAP transmitting orthogonal waveforms.
To further enhance the detection performance, there have been ever-increasing interest in jointly optimizing transmit waveforms and receive filters for MIMO-STAP \cite{tang2016joint, Tang2016WCM, tang2020polyphase,cui2017stap,2020manifold,li2022maximin}.
In \cite{tang2016joint,cui2017stap,2020manifold}, the authors considered the maximization of SINR under several practical constraints on the sought waveforms, including the constant-envelope constraint and the similarity constraint.
A number of algorithms were developed therein to tackle the joint design problems efficiently.
In \cite{tang2020polyphase}, the authors extended the algorithm in \cite{tang2016joint} to design finite-alphabet waveforms.
In \cite{Tang2016WCM,li2022maximin}, the authors focused on the robust design for MIMO-STAP under circumstance of prior knowledge mismatch. It was shown that the synthesized waveforms based on maximizing the worst-case SINR exhibited increased robustness.

Note that an operating AEW system not only detects targets from hundreds miles away, but also might communicates with friendly aircrafts/ships to perform command and control.
Therefore, if the radar and the communication systems onboard share the same frequency band, they will interfere each other.
Moreover, in a spectrally crowded environment, in which the radar has to operate with many nearby radiators simultaneously, the possibly severe mutual interference will degrade the system performance significantly.
One possible way to improve the radar performance in spectrally crowded environments is by transmitting intelligent waveforms \cite{Griffiths2015spectrum}.
In \cite{Aubry2014nonconvex,tang2018alternating,tang2019spectrally,Wu2019SpectralShaping,aubry2020design,yang2020design,Fan2020minmax,Yang2022Multispectrally}, the authors considered the waveform design under a spectral constraint.
It was shown that the spectrally constrained waveforms formed notches in the stopbands (i.e., the frequency bands that the nearby radiators operate in), thus enhancing the spectral compatibility of the radar system.

In this paper, we consider the joint design of transmit waveforms and receive filters for MIMO-STAP of AEW systems in spectrally crowded environments.
Considering that multiple nearby radiators might be present and to guarantee the quality of service of these radiators, we impose a multi-spectral constraint on the waveforms. Moreover, to minimize the distortion due to the nonlinear effects in high power amplifier, a peak-to-average-power ratio (PAPR) constraint is imposed.
We assume that the operating frequency band of the nearby radiators are known \emph{a priori} (see also similar assumptions in \cite{Aubry2014nonconvex,Wu2019SpectralShaping,aubry2020design,Fan2020minmax,Yang2022Multispectrally}).  Indeed, such prior knowledge can be obtained by cognitive methods in \cite{Aubry2018Multi-snapshot,Aubry2018High,DeMaio2017Cognition}.
Motivated by \cite{tang2016joint, tang2020polyphase}, we develop two cyclic optimization methods to jointly design the waveforms and the filters. For the challenging non-convex waveform design problem (for fixed filters), we use Dinkelbach's transform \cite{1967dinkelbach} to transform the fractional objective function into a quadratic function. Then we resort to the coordinate-descent (CD) method to split the quadratic problem into multiple subproblems, and use the alternating direction method of multipliers (ADMM) to deal with the resulting quadratically constrained quadratic programming (QCQP) problem
(we call it the DK-ADMM).
Alternatively, we also use the minorization-maximization (MM) technique to construct a quadratic surrogate of the objective, and leverage the CD and ADMM to design the transmit waveforms (we call it MM-ADMM).
We highlight that the proposed iterative algorithm in this paper can start from an infeasible point (i.e., a waveform not satisfying the constraints) and the performance of the devised waveforms is insensitive to the initial points. Moreover, the proposed algorithm can achieve better target detection performance than the competing algorithms.


The rest of this paper is organized as follows: Section II establishes the signal model and formulates the waveform design problem. Section III develops
a cyclic method to optimize the receive filters and transmit waveforms. Section IV provides numerical examples to demonstrate the performance of the proposed algorithm. Finally, conclusions are drawn in Section V.

\emph{Notations}: See Table \ref{tab1}.
\begin{table}[!htbp]
  \caption{{{List of Notations}}}
  \centering
  \begin{tabular}{cl}
   \hline
   Symbol & Meaning\\
   \hline
   $\bM$ & Matrix \\
   $\bx$ & Vector\\
   $x$ & Scalar\\
   $\bI_N$ & $N \times N$ identity matrix \\
   $(\cdot)^\ast$, $(\cdot)^\top$, $(\cdot)^\dagger$   & Conjugate, transpose, conjugate transpose\\
   $(\cdot)^{1/2}$   & Squared root of a positive semi-definite matrix\\
   $\rm{tr}(\cdot)$ & Trace of a square matrix\\
   $\left| \cdot \right|$, $\left\| \cdot \right\|_2$, $\left\| \cdot \right\|_\textrm{F}$ & Magnitude, Euclidian norm (of a vector), and \\
        & Frobenius norm (of a matrix) \\
   $\expE\{ \cdot \}$ & Expectation of a random variable\\
   $\realR$, $\complexC$ & Domain of the real and complex numbers\\
   $\rm{vec}(\cdot)$ & Vectorization\\
   $\bA \otimes \bB$ & Kronecker product\\
   $\rm{Re}(\cdot)$ & The real part of a (complex-valued) matrix\\
   $\mathcal{U}(\cdot)$ & Uniform distribution\\
   $\bA \succ \bzero$ $(\bA \succeq \bzero)$ & $\bA$ is positive definite (semi-definite)\\
   \hline
  \end{tabular}
  \label{tab1}
 \end{table}

\section{Signal Model and Problem Formulation}


\begin{figure}[!htp]
 \centering
 \includegraphics[width=0.5\textwidth]{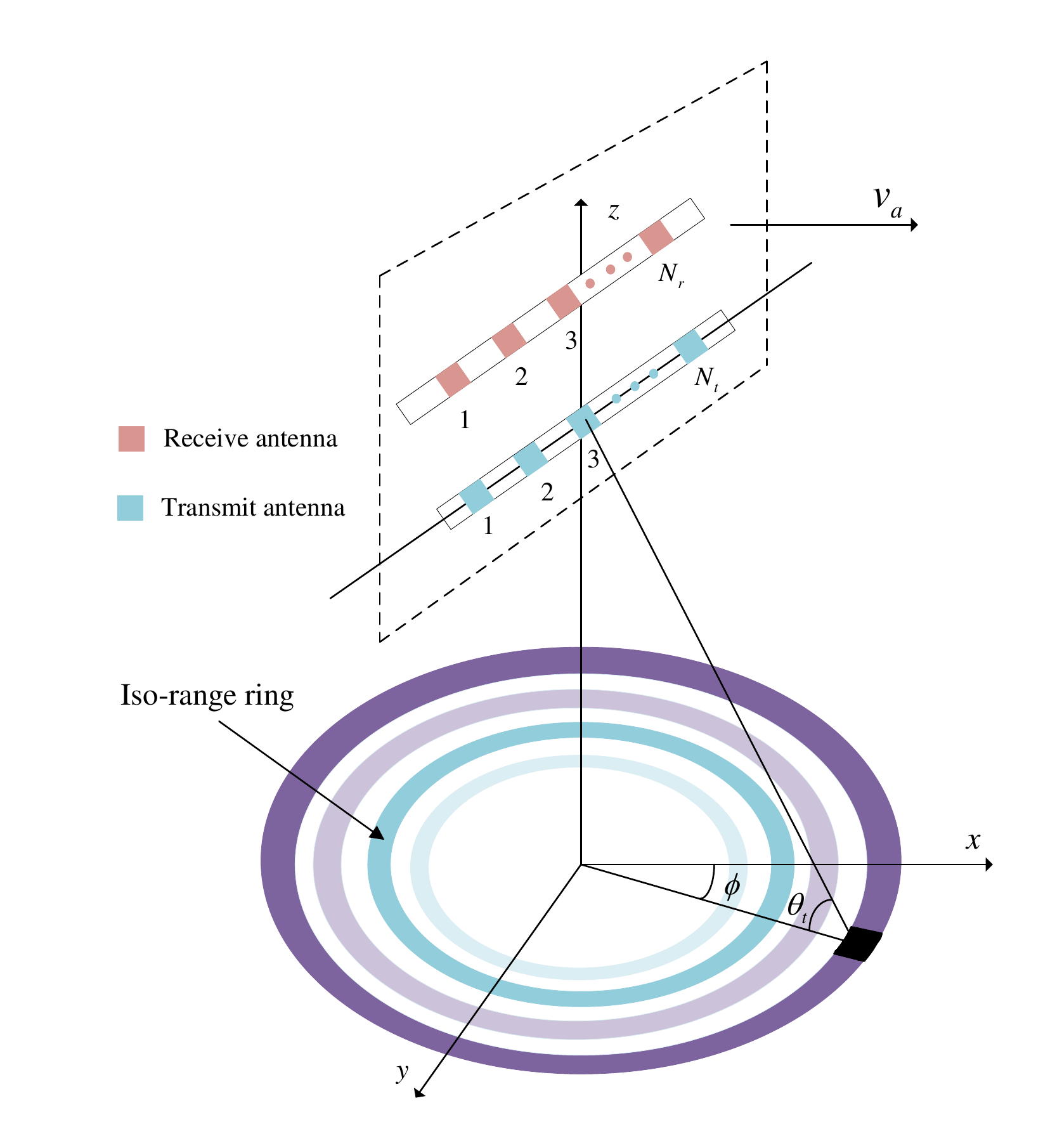}
  \caption{Geometry of an airborne MIMO STAP radar.}
  \label{Fig:Stap}
\end{figure}

\subsection{Signal Model}
As shown in \figurename ~\ref{Fig:Stap}, the considered AEW MIMO radar system has $N_t$ transmit antennas and $N_r$ receive antennas. Let $\bs_n\in\mathbb{C}^L$ be the (discrete-time) baseband waveform of the $n$th transmitter, where $L$ is the code length. Let $\bS=[\bs_1,\bs_2,\cdots,\bs_{N_t}]^\top \in\mathbb{C}^{N_t \times L}$ denote the transmit waveform matrix. Assume that the airborne MIMO radar system transmits a burst of $M$ pulses in a coherent processing interval (CPI) with the pulse repetition frequency (PRF) denoted $f_r$. For a down-looking airborne MIMO radar system, the received signal includes the target returns, the signal-dependent clutter, and the receiver noise. Next we present the signal model associated with these components (we refer to \cite{tang2016joint,tang2020polyphase} for more details).

\begin{enumerate}
\item
\textbf{Target}\\
Assume that the transmit waveforms are narrowband. Under the far-field assumption, the target return from the $m$th pulse ($m=1,2,...,M$) can be expressed as
\begin{equation}  \label{eq:Y}
  \bY_{t,m}=\alpha_te^{j(m-1)w_t}\bb(\theta_t)\ba^{\top}(\theta_t)\bS,
\end{equation}
where $\alpha_t$ is the target amplitude, $w_t=2\pi f_t$, $f_t$ is the normalized target Doppler frequency, $\theta_t$ is the target direction of arrival (DOA), $\ba(\theta_t)$ and $\bb(\theta_t)$ are the transmit array steering vector and the receive array steering vector at $\theta_t$, respectively.
Let $\by_{t,m}=\textrm{vec}(\bY_{t,m})$, $\bs=\textrm{vec}(\bS)$, and $\bA(\theta_t)=\bb(\theta_t)\ba^\top(\theta_t)$. Then
\begin{equation}  \label{eq:y}
  \by_{t,m}=\alpha_te^{j(m-1)w_t}(\bI_L\otimes\bA(\theta_t))\bs.
\end{equation}
Let $\by_t=[\by_{t,1}^\top,\cdots,\by_{t,M}^\top]^\top \in\mathbb{C}^{LMN_r}$. Then $\by_t$ can be expressed as
\begin{equation}  \label{eq:yt}
  \by_t=\alpha\bV(w_t,\theta_t)\bs,
\end{equation}
where $\bV(w_t,\theta_t)=\bd(w_t)\otimes\bI_L\otimes\bA(\theta_t)$ with $\bd(w_t)=[1,\cdots,e^{j(M-1)w_t}]^\top$ being the temporal steering vector at the Doppler frequency $f_t$.

\item
\textbf{Clutter}\\
The clutter refers to signal-dependent interference due to unwanted reflections, e.g., from ground, sea, etc. The clutter can be much stronger than the target echoes, due to the large number of clutter patches in the iso-range rings (including the range ring that the target is present and the neighborhood range rings), as shown in \figurename ~{\ref{Fig:Stap}}. Additionally, the clutter is distributed in Doppler domain owing to the motion of AEW platform \cite{ward1998space}. Assume that there are $2P+1$ clutter rings under consideration, and we split each clutter ring into $N_c$ clutter patches uniformly. Assume that the target is at the $r$th range cell, the clutter associated with the $m$th pulse, the $(r+p)$th range cell, and the $k$th patch in azimuth, can be modeled by
\begin{align}  \label{eq:Y_cpk}
  \bY_{c,m,p,k}= &\alpha_{c,p,k}e^{j2\pi(m-1)f_{c,p,k}T_r}\nonumber \\
                         &\times \bb(\theta_{c,p,k})\ba^\top(\theta_{c,p,k})\bS\bJ_p,
\end{align}
where $\alpha_{c,p,k}$, $f_{c,p,k}$, $\theta_{c,p,k}$ are the amplitude, the Doppler frequency, and the DOA of the $k$th clutter patch in the $(r+p)$th range cell, respectively, $\bJ_p=\bJ_{-p}^\top\in\mathbb{C}^{L\times L}$ is the shift matrix expressed as
\begin{equation}  \label{eq:J_p}
  \bJ_p(m,n)=
  \begin {cases}
    1, \textrm{if}\ m-n+p=0,\\
    0, \textrm{if}\ m-n+p\neq 0.
  \end{cases}
\end{equation}
Let $\by_{c,p,k}=[\textrm{vec}^\top(\bY_{c,1,p,k}^\top),\cdots,\textrm{vec}^\top(\bY_{c,M,p,k}^\top)]^\top$, then the $k$th clutter patch in the $(r+p)$th range cell can be expressed as
\begin{equation}  \label{eq:y_cpk}
  \by_{c,p,k}=\alpha_{c,p,k}\bV(w_{c,p,k},\theta_{c,p,k})\bs,
\end{equation}
where $\bV(w_{c,p,k},\theta_{c,p,k})=\bd(w_{c,p,k})\otimes\bJ_p^\top\otimes\bA(\theta_{c,p,k})$, and $w_{c,p,k} = 2\pi f_{c,p,k}$.
By considering the clutter from the nearest $2P+1$ range cells, the clutter model can be established by
\begin{equation}  \label{eq:y_c}
  \by_c=\sum_{p=-P}^{P}\sum_{k=1}^{N_c} \by_{c,p,k}.
\end{equation}
Assume that the signals associated with different clutter patches are uncorrelated. Then the clutter covariance matrix, defined by
$\bR_c(\bs)=\mathbb{E}(\by_c \by_c^\dagger)$, can be expressed as
\begin{equation}  \label{eq:R_c_ini}
  \bR_c(\bs)=\sum_{p=-P}^{P}\sum_{k=1}^{N_c}\sigma_{c,p,k}^2 \bv_{c,p,k}(\bs)\bv_{c,p,k}^\dagger(\bs),
\end{equation}
where $\sigma_{c,p,k}^2=\mathbb{E}(|\alpha_{c,p,k}|^2)$ denotes the average power of the $k$th clutter patch in the $p$th range ring, and $\bv_{c,p,k}(\bs)=\bV(w_{c,p,k},\theta_{c,p,k})\bs$.


\item
\textbf{Noise}\\
Assume that the receiver noise is white, with power of $\sigma^2$. Then the noise covariance matrix can be written as:
\begin{equation}  \label{eq:R_n}
  \bR_{\textrm{u}}=\mathbb{E}(\by_{\textrm{u}} \by_{\textrm{u}}^\dagger)=\sigma^2\bI_{LMN_r},
\end{equation}
where $\by_{\textrm{u}}$ is the vector of receiver noise.
\end{enumerate}

\subsection{Design Metric}
In radar systems, the target detection performance is closely related to the SINR. Through maximizing the output SINR, the clutter can be suppressed to a low level and then the detection performance is improved. In this paper, we aim to maximize the output SINR through jointly designing the transmit waveforms and the receive filters. Let $\bw=[\bw_1^\top,\cdots,\bw_{N_r}^\top]^\top$ denote the receive filter, with $\bw_j\in \mathbb {C}^{ML}$ representing the filter in the $j$th receiver, $j=1,\cdots,N_r$. The output SINR of the MIMO-STAP radar is defined as follows
\begin{equation}  \label{eq:SINR_ini}
  \begin{aligned}
     \textrm{SINR}(\bw,\bs)&=\frac{|\bw^\dagger \by_t|^2} {\bw^\dagger \mathbb E(\by_c \by^\dagger_c +\by_{\textrm{u}}\by_{\textrm{u}}^\dagger)\bw}\\
     &=\frac{|\alpha_t|^2 |\bw^\dagger \bv_t(\bs)|^2} {\bw^\dagger \bR_v(\bs) \bw},
  \end{aligned}
\end{equation}
where $\bv_t(\bs)=\bV(w_t,\theta_t)\bs$, and $\bR_v(\bs) = \bR_c(\bs)+\bR_{\textrm{u}}$.

\subsection{Transmit Waveform Constraints}
Now we briefly discuss the constraints that the transmit waveforms should satisfy.

\begin{enumerate}
  \item
  \textbf{Energy Constraint}\\
Since the energy of transmit waveforms is limited, the energy constraint is enforced on the sought waveforms:
\begin{equation}
  \textrm{tr}(\bS\bS^\dagger)=e_t,
\end{equation}
where $e_t$ is the total transmit energy.
Note that $\bs=\textrm{vec}(\bS)$. Then we can rewrite the energy constraint as follows
\begin{equation}  \label{eq:energy_constraint}
  \bs^\dagger\bs=e_t.
\end{equation}
Note also that practical radar systems use almost identical radio frequency amplifiers (RFA), meaning that the transmit energies across different antennas are usually uniform \cite{Stoica2007Probing}. Thus, the following uniform transmit energy constraint is included:
\begin{equation}  \label{eq:energy_uniformconstraint}
  \bs_n^\dagger \bs_n=e_t/{N_t}, n=1,\cdots,N_t.
\end{equation}

\item
  \textbf{PAPR Constraint}\\
To allow the RFA to operate in a saturated condition as well as avoid nonlinear effects, transmit waveform with low PAPR are desirable \cite{cheng2017mimo,tang2021information}. Therefore, we also impose the PAPR constraint on the waveforms, that is,
  \begin{equation}  \label{eq:PAPR_constraint}
    \bs_n^\dagger \bs_n=e_t/{N_t},\ \textrm{PAPR}(\bs_n)\leq\rho,
  \end{equation}
  where $1\leq \rho \leq L$, $n=1,\cdots,N_t$, and
  \begin{equation*}  
    \textrm{PAPR}(\bs_n)=\frac{\textrm{max}_l|s_n(l)|^2}{\frac{1}{L}\sum_{l=1}^L|s_n(l)|^2}, \ l=1,\cdots,L.
  \end{equation*}
  Particularly, if $\rho=1$, the PAPR constraint is reduced to the constant-envelope constraint:
  \begin{equation*}  
    |s_n(l)|=\sqrt{p_s},\ n=1,\cdots,N_t,\ l=1,\cdots,L,
  \end{equation*}
  where $p_s=e_t/({LN_t})$.

\item
  \textbf{Multi-Spectral Constraint}\\
  Owing to the massive increase in the number of radio devices and the limited spectrum resources, radar systems may have to share the frequency band with communication systems, which will cause mutual interference and deteriorate the performance of both systems. To improve the spetral compatibility, one possible way is to control the radar transmit waveforms to form notches in the stopbands (i.e., minimize the energy spectral density (ESD) of radar transmit waveforms in the working frequency bands of communication systems). In this respect, assume that $K_{rad}$ licensed radiators are coexisting with the MIMO radar system. Let $\Omega_k=[f_1^k,f_2^k]$ denote the normalized frequency band of the $k$th radiator, where $f_1^k$ and $f_2^k$ indicate the lower and the upper normalized frequencies, $k=1, \cdots, K_{rad}$. Note that the ESD of the $n$th waveform is written as
  \begin{equation}  \label{eq:S_k(f)}
    S_n(f)=|\bs_n^\dagger\ba(f)|^2,
  \end{equation}
where $\ba(f)=[1,e^{j2\pi f},\cdots,e^{j2\pi (L-1)f}]^\top$. Therefore, the energy of $\bs_n$ leaked on the $k$th stopband can be expressed as
\begin{equation*}  
 \int_{f_1^k}^{f_2^k}S_n(f)df=\int_{f_1^k}^{f_2^k}|\bs_n^\dagger \ba(f)|^2 df=\bs_n^\dagger\bR_I^k\bs_n,
\end{equation*}
where the $(m,l)$th element of $\bR_I^k$ is given by
\begin{equation*}  
  \bR_I^k(m,l)=
  \begin {cases}
     f_2^k-f_1^k, & m=l,\\
     \frac{e^{j2\pi f_2^k(m-l)}-e^{j2\pi f_1^k(m-l)}} {j2\pi (m-l)}, &m\neq l.
  \end {cases}
\end{equation*}
To enhance the spectral compatibility of the radar signals with the licensed radiators, the following spectral constraint is enforced on the transmit waveforms, which is given by
\begin{equation}\label{eq:spectral_constraint}
  \bs_n^\dagger \bR_I^k\bs_n \leq E_I^k,
\end{equation}
where $E_I^k$ denotes the maximum allowed interference energy of $\bs_n$ on the $k$th frequency band ($n=1, \cdots, N_t, k = 1,\cdots,K_{rad}$). Note that when the constraint in \eqref{eq:spectral_constraint} is satisfied, we can precisely control the interference energy of each waveform on every frequency band, meaning that it is possible to ensure the quality of service for each licensed radiator. In the sequel, similar to \cite{Aubry2016Multiple,yang2020design,aubry2020design,Yang2022Multispectrally}, we call the constraint in \eqref{eq:spectral_constraint} a multi-spectral constraint \footnote{We point out that the multi-spectral constraint is enforced on multiple waveforms, whereas the studies in \cite{Aubry2016Multiple,yang2020design,aubry2020design,Yang2022Multispectrally} enforce the multi-spectral constraint on a single waveform.}.

\end{enumerate}

\subsection{Problem Formulation}
By considering the constraints in \eqref{eq:energy_constraint}, \eqref{eq:PAPR_constraint}, and \eqref{eq:spectral_constraint}, we formulate the following joint design problem to maximize the output SINR of MIMO radar in spectrally crowded environments:
\begin{equation} \label{eq:P}
  \mathcal{P}
  \begin{cases}
    \begin{aligned}
      \max\limits_{\bw,\bs}\ &\textrm{SINR}(\bw,\bs)\\
       \textrm{s.t.}\ &\bs_n^\dagger \bs_n=e_t/{N_t}, \\
        &\textrm{PAPR}(\bs_n)\leq\rho,\\
        &\bs_n^\dagger \bR_I^k\bs_n \leq E_I^k,\\
        &n=1,\cdots,N_t, \ k=1,\cdots,K_{rad}.
    \end{aligned}
  \end{cases}
\end{equation}

Note that $\mathcal{P}$ is in general a non-convex problem, due to the PAPR constraint.  In the next section, we develop a cyclic method to provide high-quality solutions to the above waveform design problem.

\emph{Remark}: In the formulation of \eqref{eq:P}, we have assumed that the prior knowledge of the interference characteristics and the operating frequency bands of the licensed radiators are available. Indeed, these knowledge can be obtained via cognitive methods (see, e.g.,  \cite{DeMiao2012Waveform, Aubry2013Knowledge, DeMaio2017Cognition, Aubry2018Multi-snapshot, Aubry2018High, Aubry2018Sequential} for more details for the application of cognitive methods in radar systems).
We also highlight that if the clutter is non-stationary (e.g., due to internal clutter motion), we will assume that the normalized Doppler frequency of the $k$th clutter patch in the $p$th range ring (i.e., $f_{c,p,k}$) is uniformly distributed around the mean$\bar{f}_{c,p,k}$, that is,
\begin{equation}
  f_{c,p,k}\sim \mathcal{U}(\bar{f}_{c,p,k} - \delta_{c,p,k}/2,\bar{f}_{c,p,k} + \delta_{c,p,k}/2),
\end{equation}
where $\delta_{c,p,k}$ rules the uncertainty of clutter Doppler frequency.
In this case, the clutter covariance matrix $\bR_c(\bs)$ can calculated by the method in \cite{Aubry2013Knowledge,Lizhihui2021Robust}.

\section{Algorithm Design}
In this section, we develop cyclic optimization methods to tackle the non-convex problem in \eqref{eq:P}. For each cyclic optimization method,  two sub-problems are involved at the $(t+1)$th iteration: the optimization of receive filters for fixed transmit waveforms (i.e., $\bs^{(t)}$ is fixed) and the optimization of transmit waveforms for fixed receive filters (i.e., $\bw^{(t+1)}$  is fixed). Next we present solutions to the two subproblems. To lighten the notations, we omit the superscripts if doing so does not have a risk of confusion.

If $\bs^{(t)}$ is fixed, the receive filters can be optimized by solving the following maximization problem:
\begin{equation}
  \max\limits_{\bw}\ \frac{ |\bw^\dagger \bv_t(\bs)|^2} {\bw^\dagger \bR_v(\bs)\bw}.
\end{equation}
It can be seen that the minimum variance distortionless response (MVDR) beamformer \cite{VanTrees2002Arraybook} maximizes the objective, i.e., the solution is given by
\begin{equation}  \label{eq:w_opt}
  \bw=\bR_v^{-1}(\bs)\bv_t(\bs).
\end{equation}

To optimize $\bs_n, n=1,\cdots,N_t$ (for fixed $\bw^{(t+1)}$), we note that the SINR can be expressed as
\begin{equation}
  \begin{aligned}
    \textrm{SINR}(\bw,\bs)&=\frac{|\alpha_t|^2 |\bw^\dagger \bV(w_t,\theta_t)\bs|^2} {\bw^\dagger \bR_c(\bs) \bw + \bw^\dagger\bR_{\textrm{u}}\bw}.
  \end{aligned}
\end{equation}
In addition,
\begin{equation}
  \bw^\dagger \bR_c(\bs)\bw = \bs^\dagger \bQ\bs,
\end{equation}
where
\begin{equation}
  \bQ = \sum_{p=-P}^{P}\sum_{k=1}^{N_c} \sigma_{c,p,k}^2\bV_{c,p,k}^\dagger\bw\bw^\dagger\bV_{c,p,k},
\end{equation}
and $\bV_{c,p,k}\triangleq\bV(w_{c,p,k},\theta_{c,p,k})$.

Let
\begin{equation}
 \bD=\bV^\dagger(w_t,\theta_t) \bw\bw^\dagger \bV(w_t,\theta_t).
\end{equation}
Then, SINR can be expressed as
\begin{equation}  \label{eq:SINR}
  \textrm{SINR}(\bw,\bs)=|\alpha_t|^2\frac{\bs^\dagger \bD\bs} {\bs^\dagger \bQ\bs+\beta(\bw)},
\end{equation}
where $\beta(\bw) = \bw^\dagger\bR_{\textrm{u}}\bw$.

Therefore, the optimization of the multiple transmit waveforms (given $\bw^{(t+1)}$) can be given by
\begin{equation}
  \mathcal{P}_{\bs}
  \begin{cases}
    \begin{aligned}
       \max\limits_{\bs}\ &\frac{\bs^\dagger \bD\bs} {\bs^\dagger \bQ\bs+\beta(\bw)}\\
       \textrm{s.t.}\ &\bs_n^\dagger \bs_n=e_t/{N_t}, \\
       &\textrm{PAPR}(\bs_n)\leq\rho, \\
       &\bs_n^\dagger \bR_I^k\bs_n \leq E_I^k, \\
       &n=1,\cdots,N_t, k=1,\cdots,K_{rad}.
    \end{aligned}
  \end{cases}
  \end{equation}


Note that $\mathcal{P}_{\bs}$ is a fractional programming problem. Next we resort to the Dinkelbach's transform \cite{1967dinkelbach} and MM to replace the fractional objective with a quadratic surrogate, respectively. Then, with the quadratic surrogate function, we propose an ADMM algorithm to tackle the non-convex QCQP problem. The corresponding algorithms are referred to as DK-ADMM and MM-ADMM, respectively.

\subsection {DK-ADMM}
Let $\bs^{(t,l)}$ denote the waveform in the $(t,l)$th iteration of the proposed algorithm, where the superscript $t$ denotes the outer iteration for the cyclic optimization, and $l$ denotes the inner iteration for Dinkelbach's transform. Let $f^{(t,l)}$ denote the SINR associated with $\bs^{(t,l)}$. By applying the Dinkelbach's transform, we formulate the following optimization problem at the $(t,l+1)$th iteration
\begin{equation}
  \hat{\mathcal{P}}_{\bs}
\begin{cases}
  \begin{aligned}
     \max\limits_{\bs}\ &\bs^\dagger\hat{\bT} \bs\\
     \textrm{s.t.}\ &\bs_n^\dagger \bs_n=e_t/{N_t},\\
     &\textrm{PAPR}(\bs_n)\leq\rho,\\
     &\bs_n^\dagger \bR_I^k\bs_n \leq E_I^k, \\
     & n=1,\cdots,N_t, k=1,\cdots,K_{rad}.
  \end{aligned}
\end{cases}
\end{equation}
where $\hat{\bT}=\bT+\eta \bI$,
\begin{equation}
  \bT=\bD-f^{(t,l)}(\bQ+\beta(\bw) /e_t\cdot \bI),
\end{equation}
and $\eta$ is a constant to ensure $\hat{\bT}\succeq \bzero$.
%

Next we use the block coordinate descent (CD) method to deal with the optimization problem $\hat{\mathcal{P}}_{\bs}$ (We refer to \cite{wright2015coordinate} for a comprehensive review of the CD method). To apply the CD method, we define $\bar{\bs} = \textrm{vec}(\bS^\top)$. Note that $\bs = \bP\bar{\bs}$ \cite{tang2016joint}, where $\bP$ is a commutation matrix. Therefore, the objective function of $\hat{\mathcal{P}}_{\bs}$ can be rewritten as
\begin{equation}
  \bs^\dagger\hat{\bT} \bs = \bar{\bs}^\dagger\bar{\bT} \bar{\bs},
\end{equation}
where $\bar{\bT} = \bP^\dagger \hat{\bT} \bP$. Next, let us partition $\bar{\bT}$ into $N_t \times N_t$  blocks, each of which is an $L\times L$ matrix. Let $\bar{\bT}_{n,m}$ denote the $(n,m)$th block of $\bar{\bT}$. Then $\bar{\bs}^\dagger\bar{\bT}\bar{\bs}$ can be rewritten as
\begin{equation} \label{eq:quadraticObjectiveSub}
  \bar{\bs}^\dagger\bar{\bT}\bar{\bs} = \bs_n^\dagger \bar{\bT}_{n,n} \bs_n+2\textrm{Re}(\bs_n^\dagger\sum_{\substack{m=1\\m\neq n}}^{N_t} \bar{\bT}_{n,m} \bs_{m})+const_0,
\end{equation}
where
\begin{equation}
  const_0= \sum_{\substack{m=1\\m\neq n}}^{N_t}\sum_{\substack{m^\prime=1\\m^\prime \neq n}}^{N_t} \bs_m^\dagger \bar{\bT}_{m,m^\prime} \bs_{m^\prime}.
\end{equation}

Based on the observation in \eqref{eq:quadraticObjectiveSub}, we formulate the following problem to optimize  $\bs_n$:
\begin{equation} \label{eq:optimize_s_n_DK}
  \mathcal{P}_{s_n}
  \begin{cases}
    \begin{aligned}
       \max\limits_{\bs_n}\ &\bs_n^\dagger\bar{\bT}_{n,n}\bs_n+2\textrm{Re}(\bb_n^\dagger \bs_n) \\
       \textrm{s.t.}\  &\bs_n^\dagger \bs_n=e_t/{N_t},\\
       &\textrm{PAPR}(\bs_n)\leq\rho, \\
        &\bs_n^\dagger \bR_I^k\bs_n \leq E_I^k, \ k=1,\cdots,K_{rad},
    \end{aligned}
  \end{cases}
\end{equation}
where
\begin{equation}
  \bb_n=\sum_{\substack{m=1\\m\neq n}}^{N_t}\bar{\bT}_{n,m} \bs_{m}.
\end{equation}
Next we use the ADMM method to deal with the optimization problem $\mathcal{P}_{s_n}$ (we refer to \cite{Boyd2011ADMM} for a tutorial review of the ADMM method).  To proceed, we reformulate $\mathcal{P}_{s_n}$ as
\begin{equation} \label{eq:ADMM_problem}
  \mathcal{P}_{\bs_n,t,\bg_k,\bz}
  \begin{cases}
    \begin{aligned}
       \max\limits_{\bs_n,t,\bg_k,\bz}\ &t+2\textrm{Re}(\bb_n^\dagger \bs_n) \\
       \textrm{s.t.}\  &\bs_n^\dagger \bs_n=e_t/{N_t},\\
        &\textrm{PAPR}(\bs_n)\leq\rho, \\
        &\bg_k=\bB_k^{1/2} \bs_n,\\
        &\|\bg_k\|^2\leq 1, \ k=1,\cdots,K_{rad}, \\
        &\bz=\bar{\bT}_{n,n}^{1/2}\bs_n, \|\bz\|^2\geq t,\\
    \end{aligned}
  \end{cases}
\end{equation}
 where $t$, $\bg_k$, and $\bz$ are the introduced auxiliary variables, and $\bB_k=\bR_I^k/E_I^k$. The augmented Lagrangian function corresponding to $\mathcal{P}_{\bs_n,t,\bg_k,\bz}$ can be expressed as
\begin{equation}
  \begin{aligned}
  &L_\vartheta (\bs_n,\bz,t,\bg_k,\bc_k,\bd)\\
  =&-t-2\textrm{Re}(\bb_n^\dagger \bs_n) \\
  & +\frac{\vartheta}{2} \left\{ \sum_{k=1}^{K_{rad}}\left(||\bg_k-\bB_k^{1/2} \bs_n+\bc_k||^2-||\bc_k||^2 \right)\right\} \\
  & +\frac{\vartheta}{2}\left\{ ||\bz-\bar{\bT}_{n,n}^{1/2}\bs_n+\bd||^2-||\bd||^2 \right\},
  \end{aligned}
\end{equation}
where $\vartheta$ is the penalty parameter, $\bc_k (k=1,2,\cdots, K_{rad})$ and $\bd$ are the Lagrange multiplier vectors. 
Then, during the $(m+1)$th iteration of the ADMM method, we carry out the following steps in \eqref{eq}, shown at the bottom of this page:

Next we present solutions to \eqref{eq:update_s}, \eqref{eq:update_z,t}, and \eqref{eq:update_g}. 

\begin{figure*}[hb] 
  \hrulefill  
  \begin{subequations}
    \begin{align}
      \bs_n^{(m+1)}&=\arg\min\limits_{\bs_n}\ L_\vartheta(\bs_n,\bz^{(m)},t^{(m)},\bg_k^{(m)},\bc_k^{(m)},\bd^{(m)}), \label{eq:update_s}\\
      (\bz^{(m+1)},t^{(m+1)})&=\arg\min\limits_{\bz,t}\ L_\vartheta(\bs_n^{(m+1)},\bz,t,\bg_k^{(m)},\bc_k^{(m)},\bd^{(m)}),\label{eq:update_z,t}\\
      \bg_k^{(m+1)}&=\arg\min_{\bg_k}\ L_\vartheta(\bs_n^{(m+1)},\bz^{(m+1)},t^{(m+1)},\bg_k,\bc_k^{(m)},\bd^{(m)}),\label{eq:update_g}\\
      \bc_k^{(m+1)}&=\bc_k^{(m)}+\bg_k^{(m+1)}-\bB_k^{1/2} \bs_n^{(m+1)},\label{eq:update_c}\\
      \bd^{(m+1)}&=\bd^{(m)}+\bz^{(m+1)}-\bar{\bT}_{n,n}^{1/2}\bs_n^{(m+1)},\label{eq:update_d}
    \end{align}
    \label{eq}
  \end{subequations}
\end{figure*}

\noindent \textbf{1) Update of $\bs_n^{(m+1)}$}\\
Define
\begin{equation}
  \bY_n=-\frac{\vartheta}{2}(\bar{\bT}_{n,n}+\sum_{k=1}^{K_{rad}}\bB_k),
\end{equation}
and
\begin{equation}
  \bh=\frac{\vartheta}{2} (\bar{\bT}_{n,n}^{1/2}(\bz+\bd)+\sum_{k=1}^{K_{rad}}\bB_k^{1/2}(\bg_k+\bc_k)).
\end{equation}
Let $\bv=\bh+\bb$. Then the update of $\bs_n^{(m+1)}$ can be given by
\begin{equation}
  \mathcal{P}_{\bs_n}^{(m+1)}
  \begin{cases}
    \begin{aligned}
      \max\limits_{\bs_n}\ & \bs_n^\dagger \bY_n \bs_n+2\textrm{Re}(\bs_n^\dagger \bv)\\
      \textrm{s.t.}\  &\bs_n^\dagger \bs_n=e_t/{N_t},\\
      &\textrm{PAPR}(\bs_n)\leq\rho. \\
    \end{aligned}
  \end{cases}
\end{equation}

We can tackle the maximization problem $\mathcal{P}_{\bs_n}^{(m+1)}$ leveraging the MM method \cite{Hunter2004MM}. To proceed, note that
  \begin{equation} \label{eq:MM}
    (\bs_n-\bs_n^{(m,j)})^\dagger (\bY_n-\lambda_{\min}(\bY_n)\bI)(\bs_n-\bs_n^{(m,j)})\geq 0,
  \end{equation}
  where $\bs_n^{(m,j)}$ is the waveform at the $(m,j)$th iteration, and $\lambda_{\min}(\bY_n)$ is the smallest eigenvalue of $\bY_n$. We can derive from \eqref{eq:MM} that
  \begin{equation} \label{eq:MM2}
    \bs_n^\dagger \bY_n \bs_n \geq 2 \textrm{Re}(\bs_n^\dagger(\bY_n-\lambda_{\min}(\bY_n)\bI)\bs_n^{(m,j)})+const_1,
  \end{equation}
  where $const_1=-(\bs_n^{(m,j)})^\dagger \bY_n \bs_n^{(m,j)}+2\lambda_{\min}(\bY_n)e_t/N_t$.
  Let
  \begin{equation}\label{eq:uDef}
    \bu^{(m,j)}=(\bY_n-\lambda_{\min}(\bY_n)\bI)\bs_n^{(m,j)}+\bv,
  \end{equation}
 then the minorized problem based on \eqref{eq:MM2} at the $(m,j+1)$th iteration can be formulated as
  \begin{equation}  \label{eq:PAPR_s}
      \begin{aligned}
         \max\limits_{\bs_n}\ &\textrm{Re}(\bs_n^\dagger \bu^{(m,j)}) \\
         \textrm{s.t.}\  &\bs_n^\dagger \bs_n=e_t/{N_t},\\
         &\textrm{PAPR}(\bs_n)\leq\rho. \\
      \end{aligned}
  \end{equation}
In \cite{2005designing}, an algorithm is provided to solve the above problem. Particularly,  if $\rho=1$, this problem has a closed-form solution
  \begin{equation}  \label{eq:unimodular_s}
    s_n^{(m,j+1)}(l)=\sqrt{p_s}\textrm{exp}(j\textrm{arg}(u^{(m,j)}(l))),
  \end{equation}
  where $s_n^{(m,j+1)}(l)$ and $u^{(m,j)}(l)$ denote the $l$th element of $\bs_n^{(m,j+1)}$ and $\bu^{(m,j)}$, respectively.

\noindent\textbf{2) Update of $\bz^{(m+1)}$ and $t^{(m+1)}$}\\
    Let $\bq=\bar{\bT}_{n,n}^{1/2}\bs_n-\bd$, then the update of $\bz^{(m+1)}$ and $t^{(m+1)}$ can be given by
    \begin{equation}
      \mathcal{P}_{\bz, t}^{(m+1)}
      \begin{cases}
        \begin{aligned}
        \min\limits_{\bz, t}\  &\frac{\vartheta}{2}||\bz-\bq||^2-t \\
        \textrm{s.t.}\ &||\bz||^2 \geq t.
        \end{aligned}
      \end{cases}
    \end{equation}
It is evident that if $t = \|\bz\|^2$, the objective function achieves the smallest value.  As a result, we can obtain the solution to  $\mathcal{P}_{\bz, t}^{(m+1)}$ through solving the following unconstrained optimization:
\begin{align}
  \min_{\bz} \frac{\vartheta}{2}\|\bz-\bq\|^2-||\bz||^2.
\end{align}
Assume that $\vartheta>2$. Then the optimal solution to $\bz$ is shown as follows
\begin{equation}
  \bz = \frac{\vartheta\bq}{\vartheta-2}.
\end{equation}

\noindent \textbf{3) Update of $\bg_k^{(m+1)}$}\\
    Let $\bx_k=\bB_k^{1/2}\bs_n-\bc_k$. Then the update of $\bg_k$ can be given by
    \begin{equation}
      \mathcal{P}_{\bg_k}^{(m+1)}
      \begin{cases}
        \begin{aligned}
        \min\limits_{\bg_k}\  &||\bg_k-\bx_k||^2, \\
        \textrm{s.t.}\ &||\bg_k||^2 \leq 1.
        \end{aligned}
      \end{cases}
    \end{equation}
Obviously, the solution to $\mathcal{P}_{\bg_k}^{(m+1)}$ is given by
\begin{equation}  \label{eq:ADMM_g}
  \bg_k=
  \begin{cases}
    \begin{aligned}
    &\bx_k,\  &||\bx_k||^2\leq 1,\\
    &\bx_k/||\bx_k||,\  &||\bx_k||^2 > 1 .\\
    \end{aligned}
  \end{cases}
\end{equation}

We sum up the proposed ADMM algorithm in Algorithm \ref{Alg:1}, where the algorithm terminates if $\|\br^{(m+1)}\|<\xi$ or the algorithm reaches a maximum number of iterations, $\xi>0$ is a small user-defined value, and
\begin{align}
  \br^{(m+1)} =&\bz^{(m+1)}-\bar{\bT}_{n,n}^{(m+1)}\bs_n^{(m+1)} \nonumber\\
                           &+\sum_{k=1}^{K_{rad}}(\bg_k^{(m+1)}-\bB_k^{1/2}\bs_n^{(m+1)}).
\end{align}
\begin{algorithm}[!htbp]
  \caption{ ADMM algorithm for $\mathcal{P}_{s_n}$.} \label{Alg:1}
   \KwIn{$e_t$, $N_t$, $\bar{\bR}_{n,n}$, $\rho$, $\bR_I^k$, $E_I^k$, and $\xi$.}
   \KwOut{$\bs_n^{(t,l+1)}$.}
   \textbf{Initialize:} $m=0$, $\bs_n^{(m)}$, $\bz$, $t$, $\bg_k$, $\bc_k$, $\bd$ and $\vartheta$.\\
   \Repeat{$\|\br^{(m)}\|<\xi$}{
   \tcp*[h]{\textrm{\textit{Update of }}$\bs_n^{(m+1)}$ }\\
   $j = 0$, $\bs_n^{(m,j)} = \bs_n^{(m)}$;\\
    \Repeat{convergence}{
    Compute $\bu^{(m,j)}$ by \eqref{eq:uDef};\\
    Update $\bs_n^{(m,j+1)}$ by solving \eqref{eq:PAPR_s};\\
    $j=j+1$;\\
    }
    $\bs_n^{(m+1)} = \bs_n^{(m,j+1)}$;\\
     \tcp*[h]{\textrm{\textit{Update of }}$\bz^{(m+1)}$ }\\
    $\bq^{(m)} =\bar{\bR}_{n,n}\bs_n^{(m)} -\bd^{(m)} $;\\
    $\bz^{(m+1)}  = {\vartheta\bq^{(m)} }/{(\vartheta-2)}$;\\
    $t^{(m+1)} = \|\bz^{(m+1)}\|_2^2$;\\
      \tcp*[h]{\textrm{\textit{Update of }}$\bg_k^{(m+1)}$ }\\
      Update $\bg_k^{(m+1)}$ by \eqref{eq:ADMM_g};\\
      $\bc_k^{(m+1)}=\bc_k^{(m)}+\bg_k^{(m+1)}-\bB_k^{1/2} \bs_n^{(m+1)}$;\\
      $\bd^{(m+1)}=\bd^{(m)}+\bz^{(m+1)}-\bar{\bR}_{n,n}^{1/2}\bs_n^{(m+1)}$;\\
     $m=m+1$
   }
   $\bs_n^{(t,l+1)}=\bs_n^{(m+1)}$.
\end{algorithm}
\subsection{MM-ADMM}
Substituting \eqref{eq:w_opt} into \eqref{eq:SINR_ini},  we rewrite $\textrm{SINR}$ as
\begin{equation} \label{eq:SINR_s}
  \textrm{SINR}(\bs)=|\alpha_t|^2 \bv_t^\dagger(\bs) \bR_v^{-1}(\bs) \bv_t(\bs),
\end{equation}
According to \cite[Lemma 1]{Tang2021RangeProfiling}, $\textrm{SINR}(\bs)$ is minorized by:
\begin{equation}
  -\bs^\dagger \bR \bs + 2\textrm{Re}(\bc^\dagger \bs) + const_2,
\end{equation}
where $\bc = \bV^\dagger(w_t,\theta_t) \bR_v^{-1}(\bs^{(k)}) \bV(w_t,\theta_t) \bs^{(k)}$, $const_2 = -\textrm{tr}(\bB_k\bR_u)$,
\begin{equation}
  \begin{aligned}
    \bR = &\sum_{p=-P}^{P}\sum_{k=1}^{N_c}\sigma_{c,p,k}^2\bV_{c,p,k} \bB_k \bV_{c,p,k}^\dagger, \nonumber
  \end{aligned}
\end{equation}
and $\bB_k=\bu_k\bu_k^\dagger$, $\bu_k=\bR_v^{-1}(\bs^{(k)})\bV(w_t,\theta_t)\bs^{(k)}$, the superscript ``$k$" denotes the $k$th iteration in the MM-based algorithm.
Let $\hat{\bR}=\mu\bI-\bR$, where $\mu$ is set to ensure $\hat{\bR}>0$. By omitting the constant terms, the optimization of $\bs$ can be formulated as
\begin{equation}
  \overline{\mathcal{P}}_{\bs}
  \begin{cases}
    \begin{aligned}
       \max\limits_{\bs}\ &\bs^\dagger \hat{\bR} \bs + 2\textrm{Re}(\bc^\dagger \bs)\\
       \textrm{s.t.}\ &\bs_n^\dagger \bs_n=e_t/{N_t}, \\
       &\textrm{PAPR}(\bs_n)\leq\rho, \\
       &\bs_n^\dagger \bR_I^k\bs_n \leq E_I^k, \\
       &n=1,\cdots,N_t, k=1,\cdots,K_{rad}.
    \end{aligned}
  \end{cases}
\end{equation}
Let $\bc = \bP\bar{\bc}$, and the objective function of $\overline{\mathcal{P}}_{\bs}$ can be rewritten as
\begin{equation}
  \bs^\dagger\hat{\bR} \bs + 2\textrm{Re}(\bc^\dagger \bs) = \bar{\bs}^\dagger\bar{\bR} \bar{\bs} + 2\textrm{Re}(\bar{\bc}^\dagger \bar{\bs}),
\end{equation}
where $\bar{\bR}=\bP^\dagger \hat{\bR} \bP$. Next, let us partition $\bar{\bR}$ and $\bar{\bc}$ into $N_t \times N_t$ and $N_t \times 1$ blocks, each of which are an $L \times L$ matrix and an $L \times 1$ vector, respectively. Let $\bar{\bR}_{n,m}$ and $\bc_n$ denote the $(n,m)$th and the $n$th block of $\bar{\bR}$ and $\bar{\bc}$. Then $\bar{\bs}^\dagger\bar{\bR} \bar{\bs} + 2\textrm{Re}(\bar{\bc}^\dagger \bar{\bs})$ can be rewritten as
\begin{equation} \label{eq:block_MM}
  \bar{\bs}^\dagger\bar{\bR} \bar{\bs} + 2\textrm{Re}(\bar{\bc}^\dagger \bar{\bs}) = \bs_n^\dagger \bar{\bR}_{n,n} \bs_n + 2\textrm{Re}(\boldf_n^\dagger \bs_n) + const_3
\end{equation}
where $\boldf_n = \frac{1}{2}\sum_{\substack{m=1\\m\neq n}}^{N_t} \bar{\bR}_{n,m} \bs_{m} + \bc_n$,
\begin{equation*}
  const_3 = \sum_{\substack{m=1\\m\neq n}}^{N_t}\sum_{\substack{m^\prime=1\\m^\prime \neq n}}^{N_t} \bs_m^\dagger \bar{\bR}_{m,m^\prime} \bs_{m^\prime} + \sum_{\substack{m=1\\m\neq n}}^{N_t} \bc_m^\dagger \bs_m.
\end{equation*}
By using \eqref{eq:block_MM}, we formulate the following problem to optimize $\bs_n$:
\begin{equation} \label{eq:optimize_s_n}
  \overline{\mathcal{P}}_{s_n}
  \begin{cases}
    \begin{aligned}
       \max\limits_{\bs_n}\ &\bs_n^\dagger\bar{\bR}_{n,n}\bs_n+2\textrm{Re}(\boldf_n^\dagger \bs_n) \\
       \textrm{s.t.}\  &\bs_n^\dagger \bs_n=e_t/{N_t},\\
       &\textrm{PAPR}(\bs_n)\leq\rho, \\
        &\bs_n^\dagger \bR_I^k\bs_n \leq E_I^k, \ k=1,\cdots,K_{rad}.
    \end{aligned}
  \end{cases}
\end{equation}
Note the similarity between \eqref{eq:optimize_s_n} and \eqref{eq:optimize_s_n_DK}. Thus, we can use Algorithm \ref{Alg:1} to tackle \eqref{eq:optimize_s_n}.

\begin{table*}[!htbp]
  \caption{{{Computational complexity analysis}}}
  \centering
  \begin{tabular}{c | c | c | c}
   \hline
   \multicolumn{2}{c|}{Algorithm 1} & \multicolumn{2}{c}{Algorithm 2} \cr\cline{1-4}
   \hline
   Computation & Complexity & Computation & Complexity \cr
   \hline
   $\bu^{(m,j)}$ & $O(L^2)$ & $\bR_c(\bs)$ & $O((2P+1)(MN_r)^3LN_t(N_t+L))$ \\
   $\bs_n^{(m,j+1)}$ & $O(L)$ & $\bw$ & $O((LMN_r)^3)$ \\
   $\bq^{(m)}$ & $O(L^2)$ & $\bD$ & $O((LN_t)^2)$ \\
   $\bz^{(m+1)}$ & $O(L)$ & $\bQ$ & $O((2P+1)(LN_t^3M^2N_r^2+L^2N_t^3MN_r))$ \\
   $t^{(m+1)}$  & $O(L)$ & $\beta(\bw^{(t+1)})$ & $O(LMN_r)$ \\
   $\bg_k^{(m+1)}$ & - or $O(L)$ & $f^{(t,l)}$& $O((LN_t)^2)$ \\
   $\bc_k^{(m+1)}$ & $O(L^2)$ & $\bar{\bT}_{n,n}^{(t,l)}$ or $\bar{\bR}_{n,n}^{(t)}$ & $O((LN_t)^2)$ \\
   $\bd^{(m+1)}$ & $O(L^2)$ & $\bb_n^{(t,l)}$ or $\boldf_n^{(t)}$ & $O((N_t-1)L^2)$ \\
  -&-& $\bB_k$ or $\bR$& $O(L^3M^2N_r^2N_t+L^2M^2N_r^2N_t)$ \\
   -& - & $\bc$ & $O(L^3M^2N_r^2N_t+L^2N_t^2)$ \\
   \hline
   \end{tabular}
   \label{Table:2}
\end{table*}

 \subsection{Algorithm Summary and Computational Complexity Analysis}
We summarize the proposed multi-spectrally constrained waveform design algorithm in Algorithm \ref{Alg:2}, where $\epsilon_1, \epsilon_2 >0$ are user-defined small values.
The computational complexity of the proposed algorithm at each iteration is analyzed in Table \ref{Table:2}. To alleviate the computational burden, we calculate $\bR_c(\bs)$ and $\bQ$ by the method in \cite[Appendix C]{tang2020polyphase}. We can see that if $P$ and $N_t$ is large, the computational complexity of DK-ADMM is higher than that of MM-ADMM; otherwise, if $L$ is large, the computational complexity of MM-ADMM is higher than that of DK-ADMM.


\begin{algorithm}[!htbp]
  \caption{ Multi-spectrally constrained waveform design for MIMO STAP.} \label{Alg:2}
   \KwIn{$\bR_{\textrm{u}}$, $\bV(w_t,\theta_t)$, $\mu$, $\epsilon_1$, $\epsilon_2$, }
   \KwOut{$\bs_{\textrm{opt}}$ and $\bw_{\textrm{opt}}$.}
   \textbf{Initialize:} $t=0$, $\bs_n^{(t)}, n=1,\cdots,N_t$.\\

   \Repeat{$|\textrm{SINR}^{(t+1)}-\textrm{SINR}^{(t)}|/\textrm{SINR}^{(t+1)}<\epsilon_2$}{
    \tcp*[h]{\textrm{\textit{Update of }}$\bw^{(t+1)}$ }\\
    Compute $\bR_c(\bs^{(t)})$ by \eqref{eq:R_c_ini};\\
    $\bR_v(\bs^{(t)})=\bR_c(\bs^{(t)})+\bR_{\textrm{u}}$;\\
    $\bw^{(t)}=\bR^{-1}_v(\bs^{(t)})\bv_t(\bs^{(t)})$; \\
    \tcp*[h]{\textrm{\textit{Update of }}$\bs^{(t+1)}$ }\\

    \For{DK-ADMM Algorithm}{
      $l = 0$, $\bs^{(t,l)} = \bs^{(t)}$;\\
      \Repeat{$|f^{(t,l+1)}-f^{(t,l)}|/f^{(t,l+1)}<\epsilon_1$}{
      Compute $\bD$, $\bQ$, and $\beta(\bw^{(t+1)})$;\\
      Compute $f^{(t,l)}$;\\
      Compute $\bar{\bT}_{n,n}^{(t,l)}$ and $\bb_n^{(t,l)}$;\\
        \For {$n=1$ $\textbf{\textrm{to}}$ $N_t$}{
          Update $\bs_n^{(t,l+1)}$ using Algorithm 1;\\
        }
        $l=l+1$;
     }
      $\bs_n^{(t+1)}=\bs_n^{(t,l)}$;
      }

    \For{MM-ADMM Algorithm}{
    Compute $\bB_k^{(t)}$, $\bR^{(t)}$, and $\bc^{(t)}$;\\
    Compute $\bar{\bR}_{n,n}^{(t)}$ and $\boldf_n^{(t)}$;\\
      \For {$n=1$ $\textbf{\textrm{to}}$ $N_t$}{
        Update $\bs_n^{(t+1)}$ using Algorithm 1;\\
      }
    }

   $t=t+1$;
   }
    $\bs_{\textrm{opt}} = \bs^{(t+1)}$;\\
    $\bw_{\textrm{opt}} = \bw^{(t+1)}$.
\end{algorithm}

\subsection{Extension to Multiple Space-Frequency Constraints}
In some situations, the directions of the licensed radiators might be approximately known. Assume that the direction of the $k$th licensed radiator belongs to $\Theta_k=[\theta_1^k,\theta_2^k]$, where $\theta_1^k$ and $\theta_2^k$ are the lower and upper angles, respectively, $k=1,\cdots,K_{rad}$. Therefore, the energy of $\bs$ leaked on the $k$th space-frequency band can be expressed as
\begin{equation*}
  \int_{v_1^k}^{v_2^k}\int_{f_1^k}^{f_2^k} |\bs_\theta^\dagger \ba(f)|^2 dfd\theta = \int_{v_1^k}^{v_2^k}\bs_\theta^\dagger \bR_I^k \bs_\theta d\theta = \bs^\dagger \bF_I^k \bs,
\end{equation*}
where $v_1^k=\sin(\theta_1^k)$, $v_2^k=\sin(\theta_2^k)$, $\bs_\theta = \textrm{vec}(\ba^\top(\theta)\bS) = (\bI_L \otimes \ba^\top (\theta))\bs$,
$\bF_I^k=\bR_I^k \otimes \bU$, the $(p,q)$th entry of $\bU \in \complexC^{N_t \times N_t}$ is given by
\begin{equation*}
  \bU(p,q)=
  \begin {cases}
     v_2^k-v_1^k, & p=q,\\
     \frac{e^{j2\pi v_2^k(q-p)d_t/\lambda}-e^{j2\pi v_1^k(q-p)d_t/\lambda}} {j2\pi (q-p)d_t/\lambda}, &p\neq q,
  \end {cases}
\end{equation*}
and we have assumed that the transmit array is a uniform linear array (ULA) with inter-element spacing denoted $d_t$.  Then we can enforce a space-frequency constraint to control the energy leaked on the space-frequency band.
When multiple space-frequency constraints and the PAPR constraint are imposed, the optimization of $\bs$ (at each iteration) can be formulated by the following:
\begin{equation} \label{eq:Problem_space_frequency}
  \mathcal{P}_{\bs}
\begin{cases}
  \begin{aligned}
     \max\limits_{\bs}\ &\frac{\bs^\dagger \bD\bs} {\bs^\dagger \bQ\bs+\beta(\bw)}\\
     \textrm{s.t.}\ &\bs^\dagger \bs=e_t,\\
     &\textrm{PAPR}(\bs)\leq\rho,\\
     &\bs^\dagger \bF_I^k\bs \leq E_I^k, \\
     & k=1,\cdots,K_{rad}.
  \end{aligned}
\end{cases}
\end{equation}
Similarly, we can use Algorithm \ref{Alg:2} to tackle the above optimization problem.



\section{Numerical Examples}
In this section, numerical experiments are conducted to evaluate the performance of the proposed algorithm.
The considered MIMO radar system has $N_t=4$ transmitters and $N_r=4$ receivers, where both transmit array and receive array are assumed to be ULAs, with inter-element spacing $d_t=2\lambda$ and $d_r=\lambda/2$, respectively ($\lambda$ is the wavelength).
The radar system is at an altitude of $h_a=9000$ m and moving with a constant speed of $v_a=75$ m/s.
The total transmit energy of the waveforms is $e_t=1$. The waveform has a bandwidth of $800$ kHz and a duration of $T=200 \mu$s, sampled with a frequency of $f_s=800$ kHz (i.e., the code length is $L=160$). Additionally, we use a linear frequency modulated (LFM) waveform with a chirp rate of $\gamma_s=3.5\times 10^9$ $\textrm{s}^{-2}$ as the initial waveform for all the transmit waveforms (Note that such waveforms do not satisfy the multi-spectral constraint, meaning that the initial waveforms are infeasible).
The radar transmits $M=16$ pulses in a CPI with a constant PRF of $f_r=1000$ Hz. The target of interest is at an azimuth of $0^\circ$, and a range of $R_t=12728$ m. To establish the clutter model, we assume that $P=3$ and $N_c=361$ clutter patches are uniformly distributed in each iso-range ring. Additionally, $\sigma_{c,p,k}^2=1, p=-P,\cdots,P,k=1,\cdots,N_c$.  The noise power is $\sigma^2=1$. $K_{rad}=3$ licensed radiators are coexisting with the AEW radar system. The normalized frequency bands of the licensed radiators are $\Omega_1=[0.2218, 0.2773]$, $\Omega_2=[0.4609,0.6132]$, and $\Omega_3=[0.7223, 0.76328]$. The maximum allowed interfered energy of each waveform on these bands are $E_I^1=-35$ dB, $E_I^2=-35$ dB, and $E_I^3=-30$ dB, respectively.
Regarding the ADMM algorithm, we set the penalty parameter to $\vartheta=4$, and the maximum number of iterations to $1000$.
For the stopping criterion of the ADMM algorithm, the Dinkelbach's transform, and the cyclic optimization, we set $\xi=5\times 10^{-10}$, $\epsilon_1=3\times 10^{-3}$, and $\epsilon_2=3\times 10^{-4}$, respectively. Finally, the experiments are conducted on a standard PC with Intel(R) Core(TM) i7-9750H CPU and 16GB RAM.


\begin{figure} [!htbp]
  \centering
  \includegraphics[width = 0.45\textwidth]{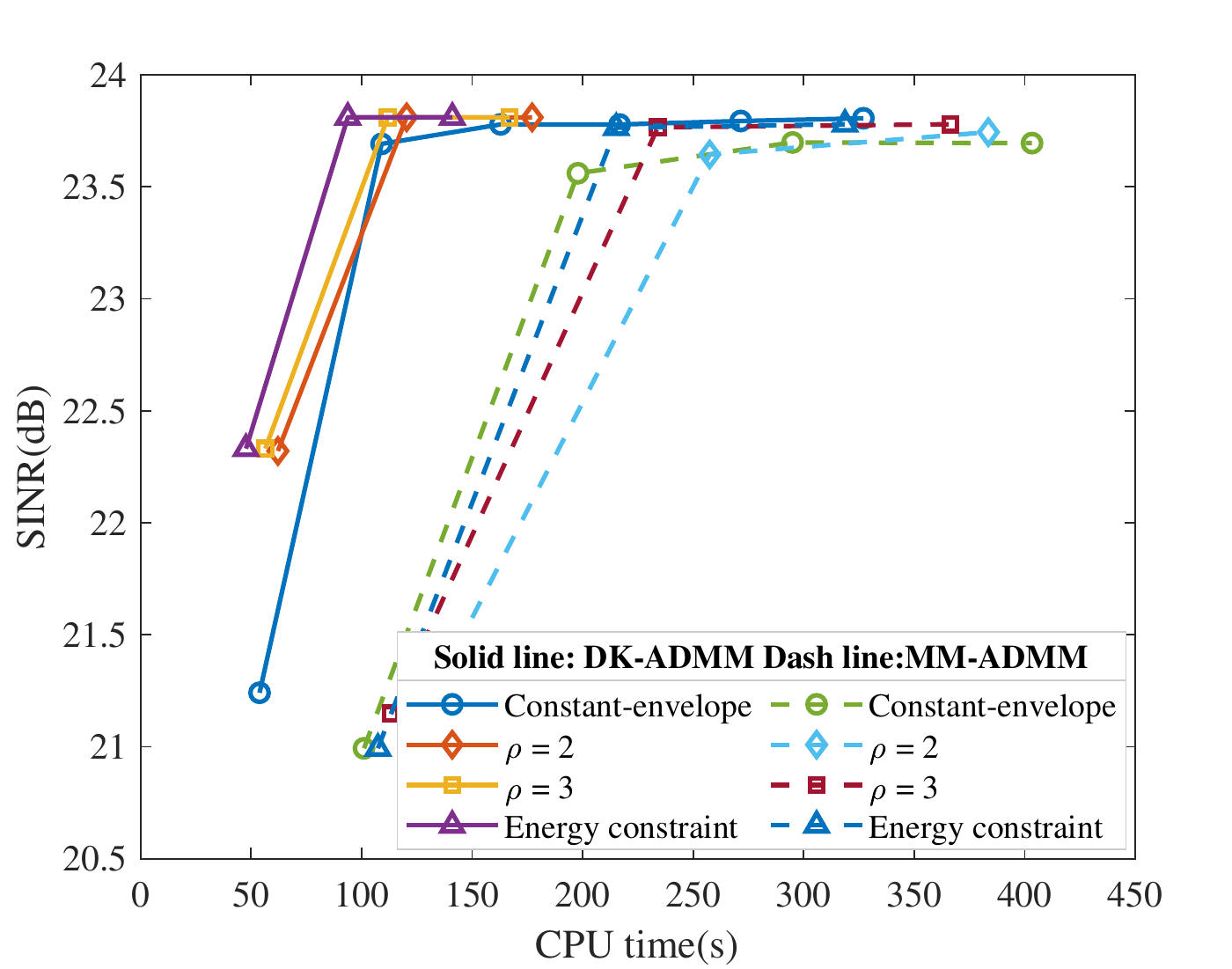}
  \caption{Convergence of SINR versus CPU time. $e_t = 1$. $v_t=52.5$ m/s. $E_I^1=E_I^2=-35$ dB, $E_I^3=-30$ dB. }
  \label{Fig:convergence}
\end{figure}

\begin{figure*}[!htbp]
  \centering
  \subfigure[$\rho =1$, DK-ADMM]{\includegraphics[width = 0.24\textwidth]{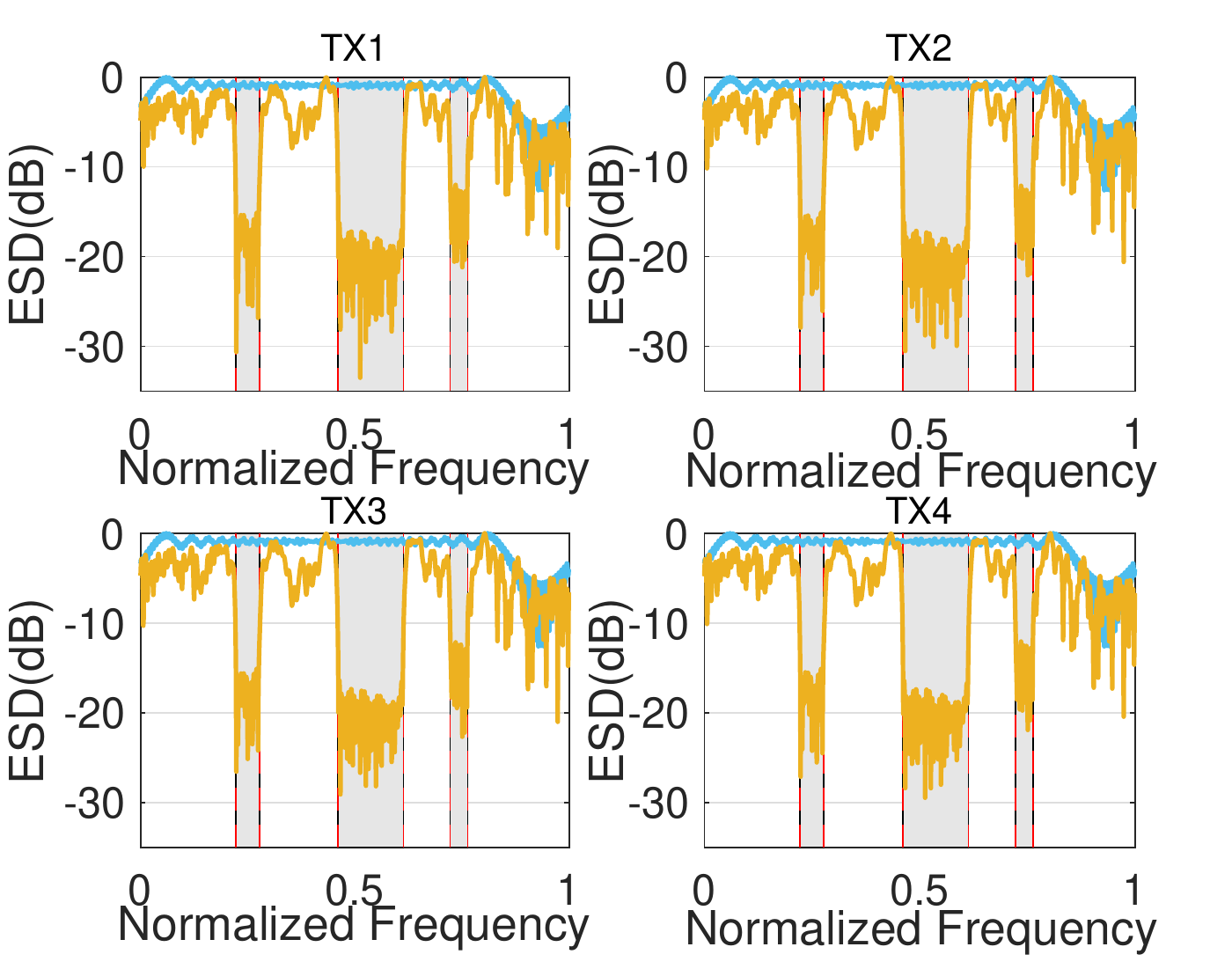}}
  \subfigure[$\rho =2$, DK-ADMM]{\includegraphics[width = 0.24\textwidth]{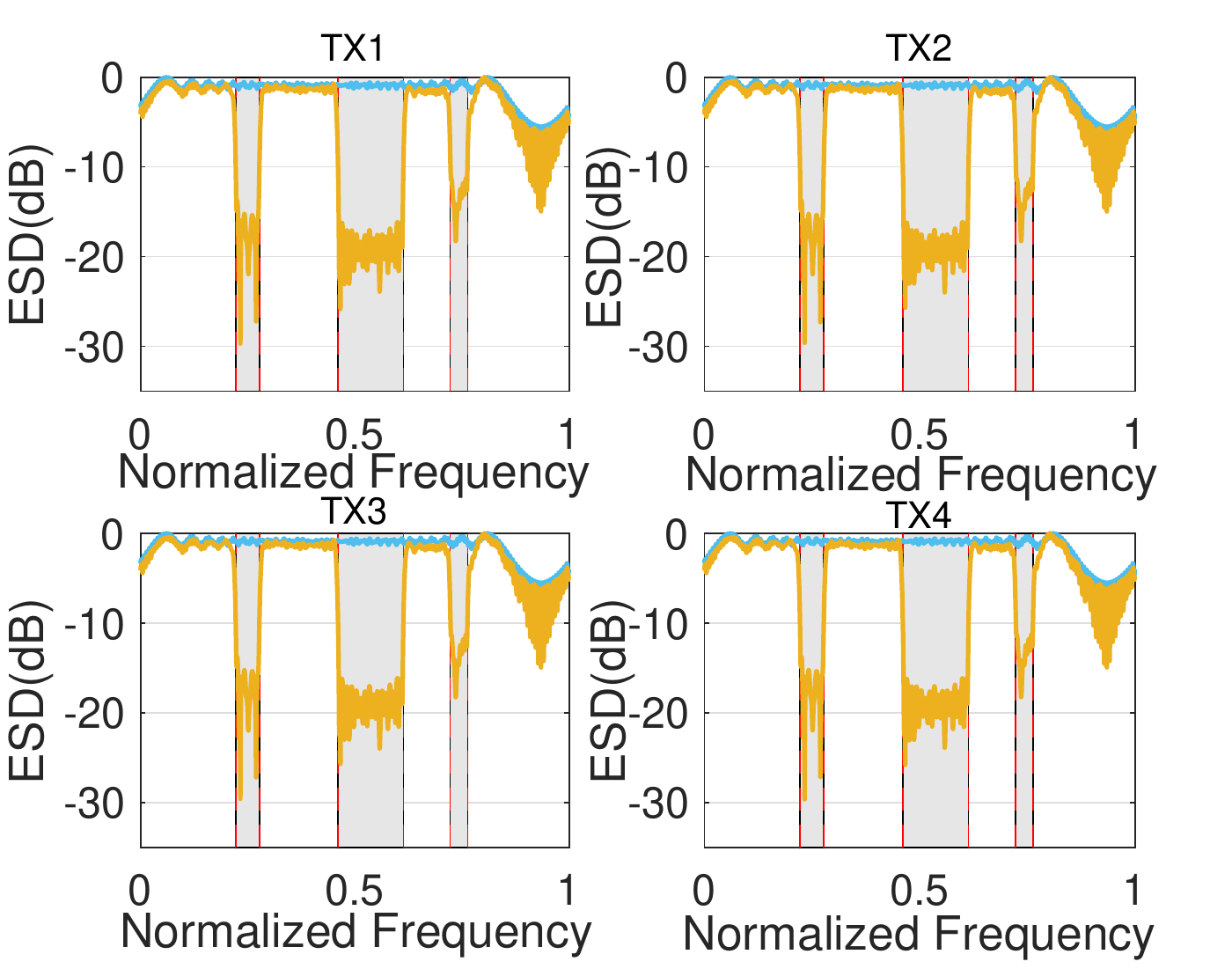}}
  \subfigure[$\rho =3$, DK-ADMM]{\includegraphics[width = 0.24\textwidth]{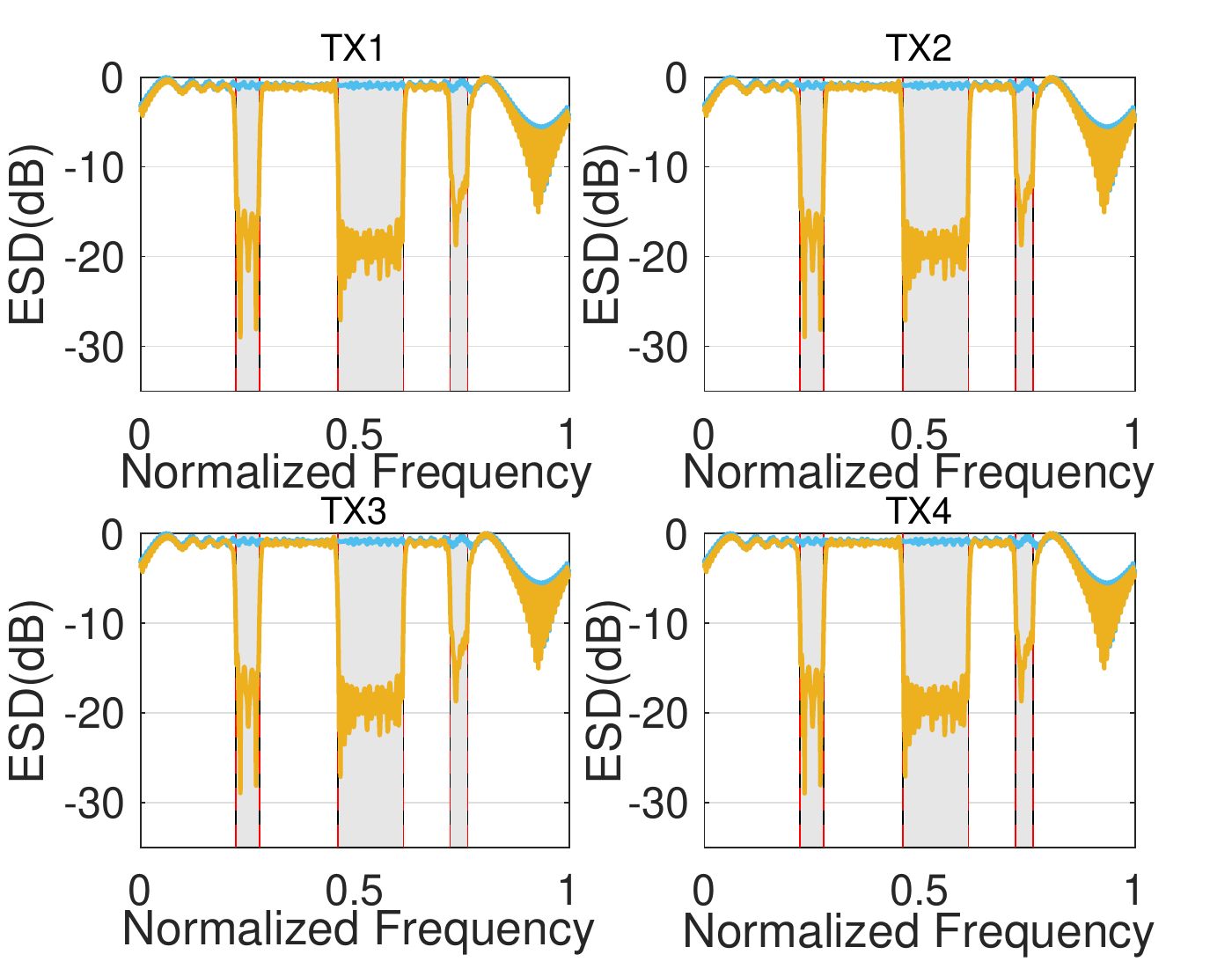}}
  \subfigure[$\rho =L$, DK-ADMM]{\includegraphics[width = 0.24\textwidth]{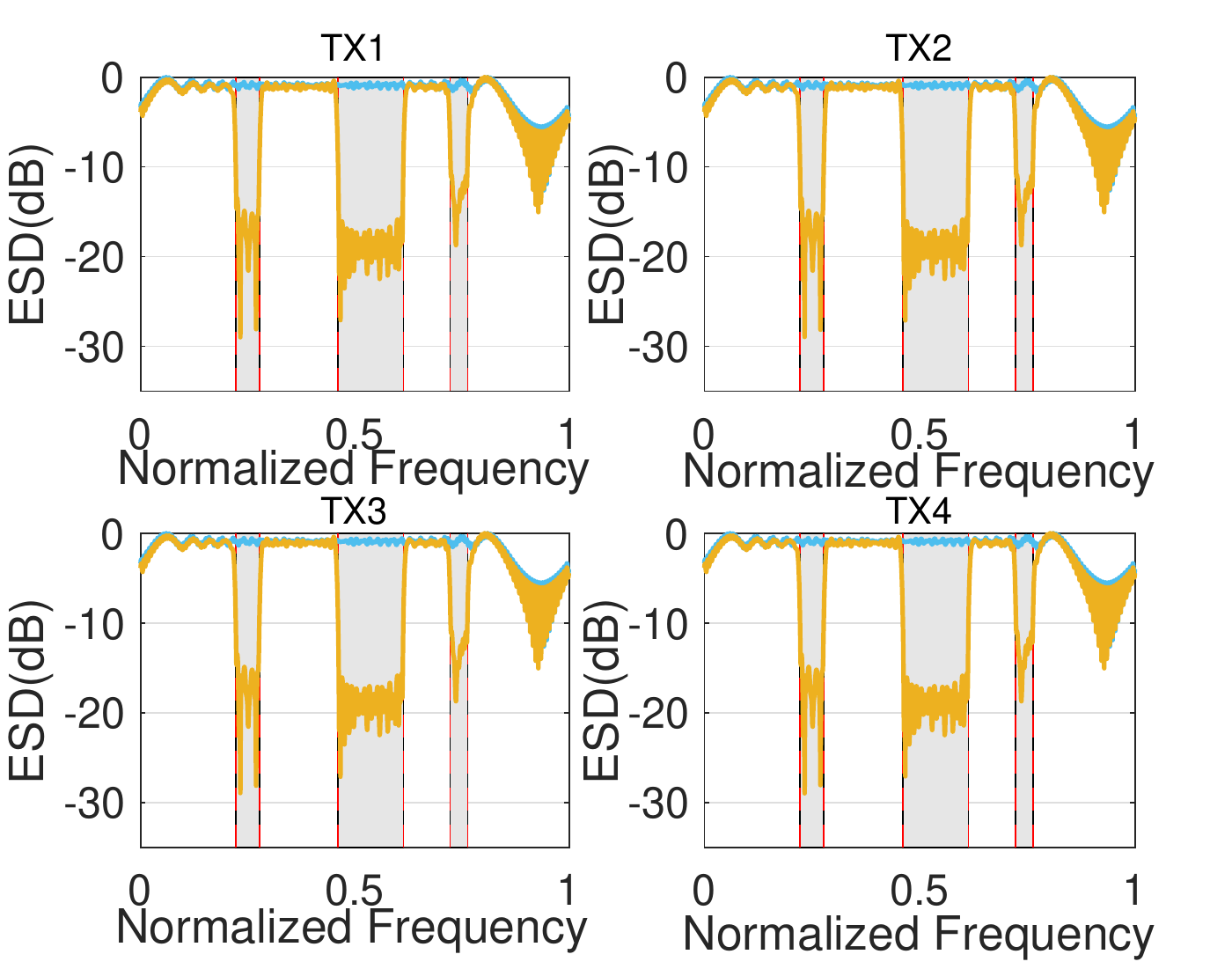}}
  \\ 
  \centering
  \subfigure[$\rho =1$, MM-ADMM]{\includegraphics[width = 0.24\textwidth]{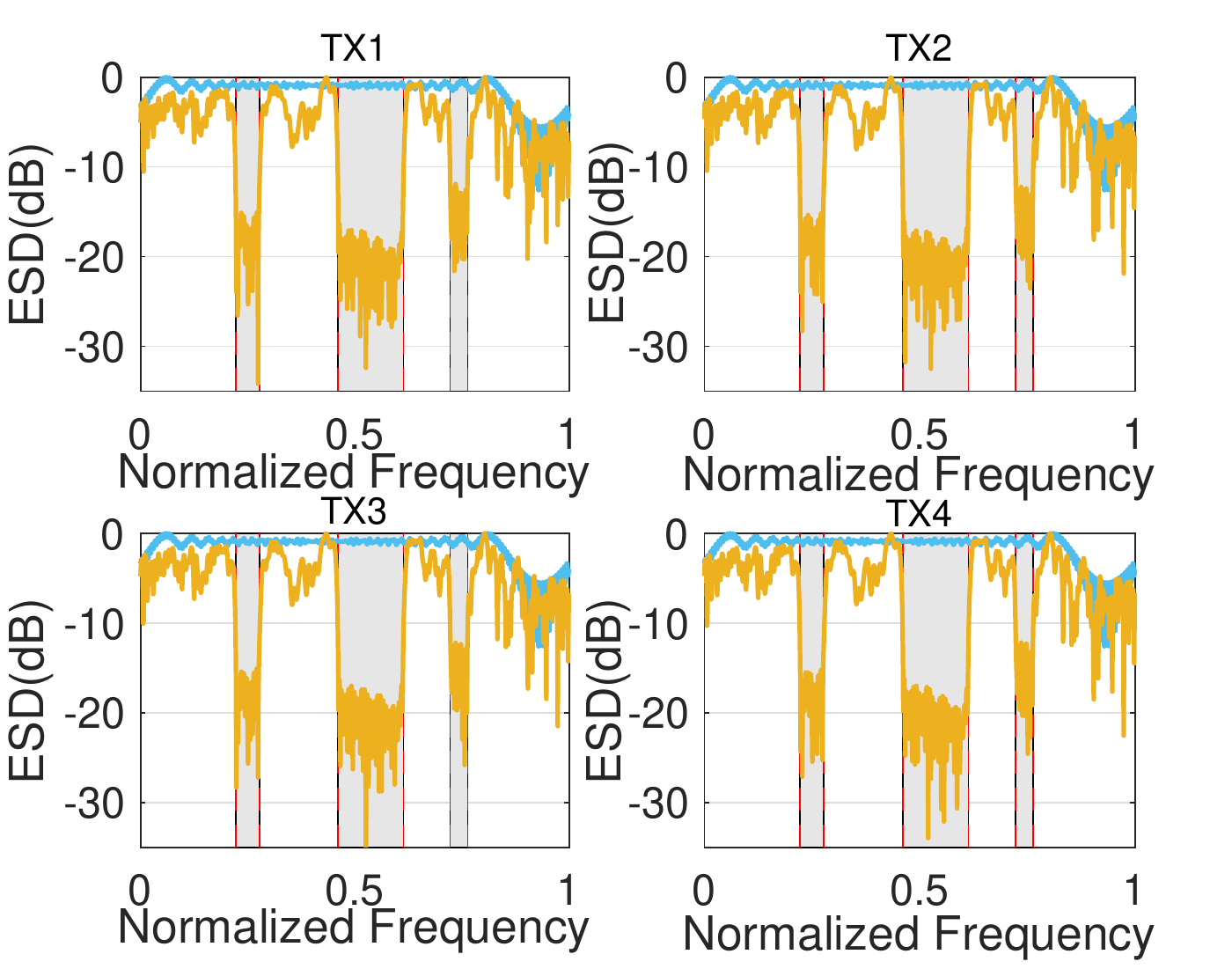}}
  \subfigure[$\rho =2$, MM-ADMM]{\includegraphics[width = 0.24\textwidth]{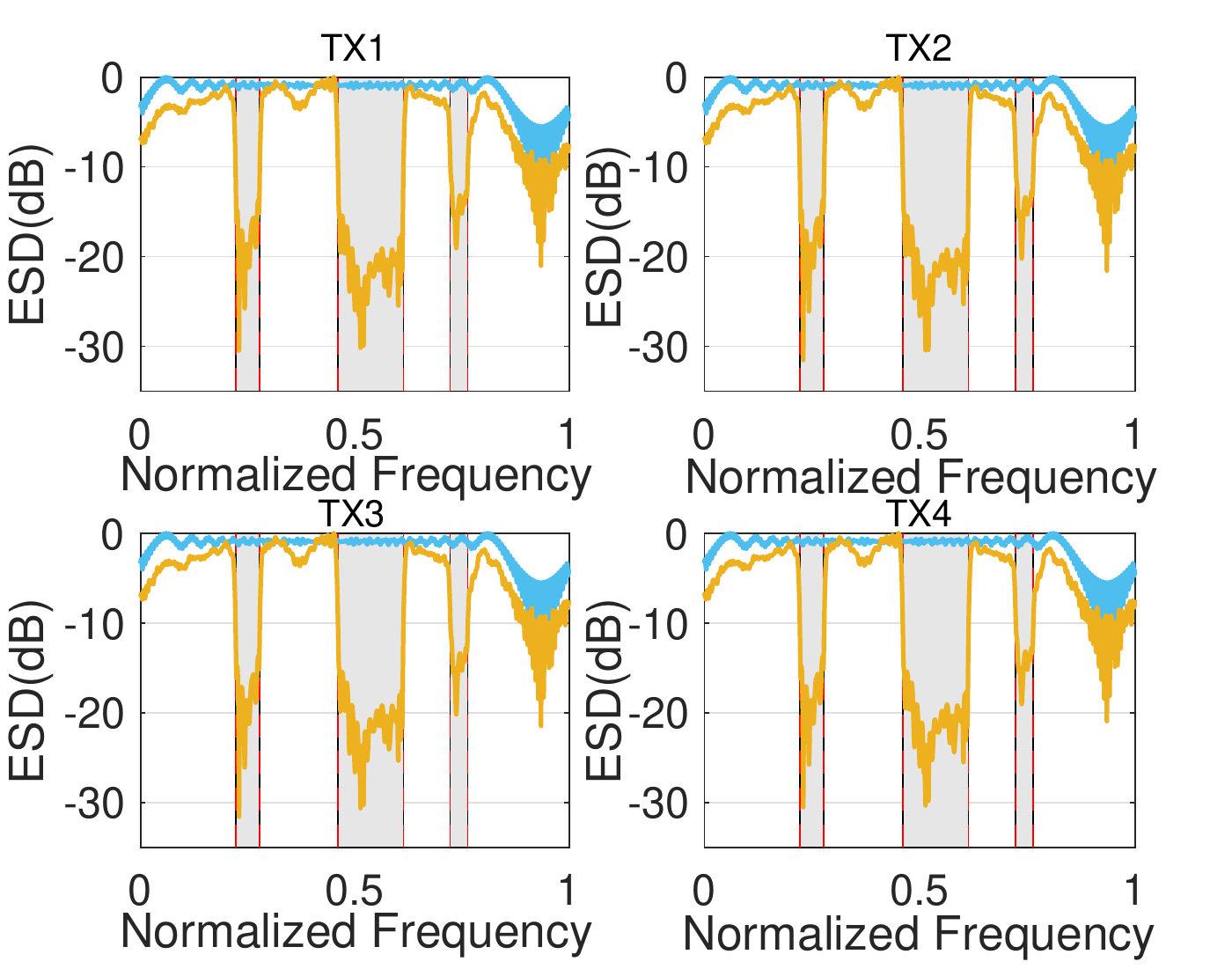}}
  \subfigure[$\rho =3$, MM-ADMM]{\includegraphics[width = 0.24\textwidth]{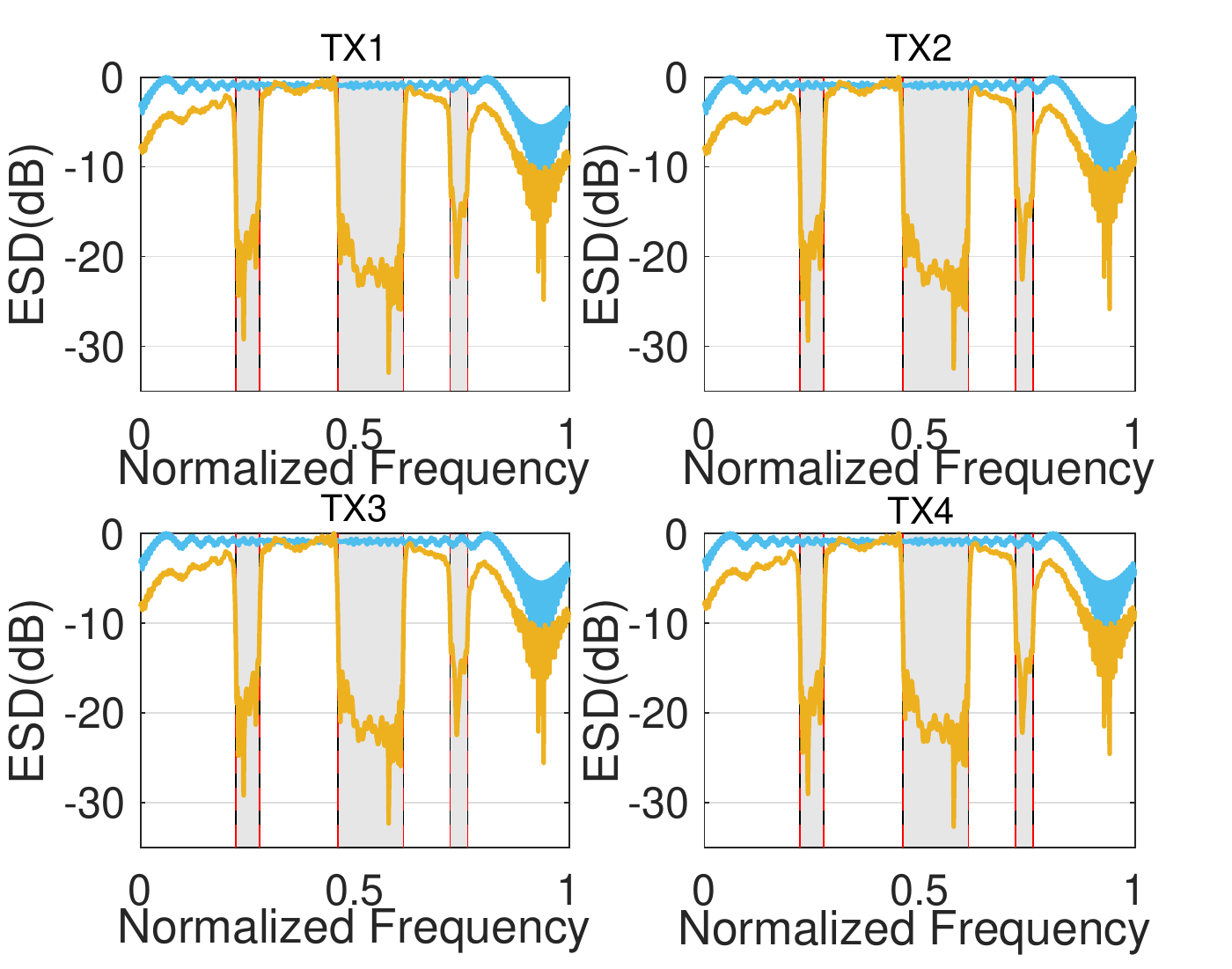}}
  \subfigure[$\rho =L$, MM-ADMM]{\includegraphics[width = 0.24\textwidth]{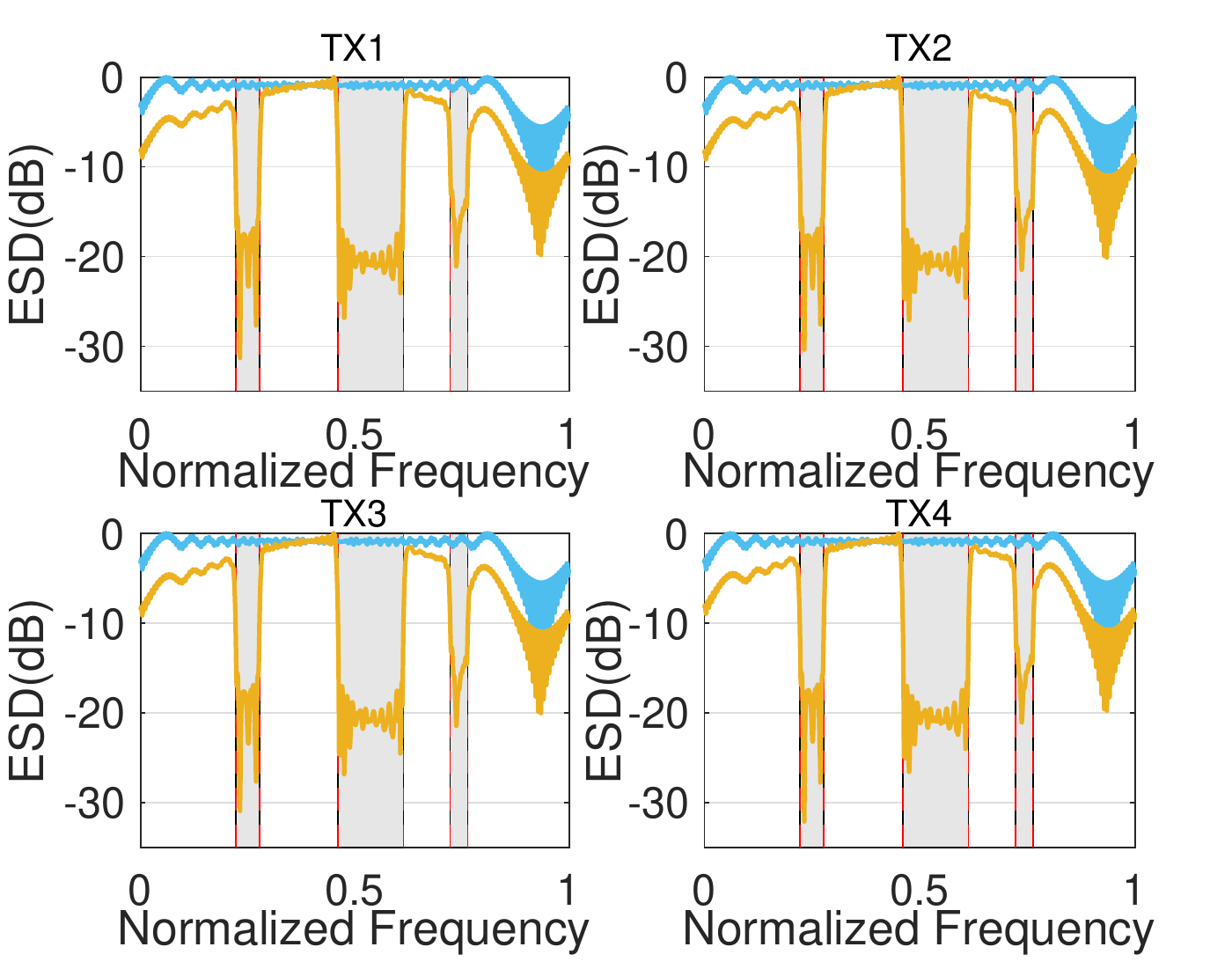}}
  \caption{ESDs of the designed waveforms. The blue and the yellow lines represent the ESDs of the initial waveforms and the optimized waveforms. $e_t=1$. $E_I^1=E_I^2=-35$ dB, $E_I^3=-30$ dB. (a) and (e): Constant-envelope waveforms. (b) and (f): $\rho=2$. (c) and (g): $\rho=3$. (d) and (h): Energy-constrained waveforms.}
  \label{Fig:ESDs}
\end{figure*}

\begin{table}[!htbp]
  \caption{{{SINR at convergence}}}
  \centering
  \begin{tabular}{c c c c c}
   \hline
    SINR (dB) & $\rho=1$ & $\rho=2$ & $\rho=3$ & $\rho=L$ \\
   \hline
   DK-ADMM & 23.8059 & 23.8101 & 23.8101 & 23.8102 \\
   MM-ADMM & 23.6952 & 23.7439 & 23.7794 & 23.7795  \\
   \hline
   \end{tabular}
   \label{Table:3}
\end{table}

\begin{table}[!htbp]
  \caption{{{CPU time needed to reach convergence}}}
  \centering
  \begin{tabular}{c c c c c}
   \hline
    CPU time (s) & $\rho=1$ & $\rho=2$ & $\rho=3$ & $\rho=L$ \\
   \hline
   DK-ADMM & 380.843 & 177.167 & 166.941 & 141.049 \\
   MM-ADMM & 403.326 & 383.573 & 366.424 & 318.698 \\
   \hline
   \end{tabular}
   \label{Table:4}
\end{table}

First, we analyze the convergence of the proposed algorithm. \figurename ~\ref{Fig:convergence} shows the SINR curves of the proposed algorithm versus the CPU time, under the PAPR constraints of $\rho=1$ (i.e., the constant-envelope constraint), $\rho=2$, $\rho=3$, and $\rho = L$ (i.e., the energy constraint), respectively, where the target velocity is $v_t=52.5$ m/s (i.e., $f_t=0.35$).
Note that for both DK-ADMM and MM-ADMM, the SINR monotonically increases as the iterations, which confirms the convergence of the proposed algorithm. The SINR of the waveforms synthesized by the DK-ADMM algorithm and the MM-ADMM algorithm at convergence is shown in Table \ref{Table:3}. We can see that a larger PAPR corresponds to a higher SINR, because of the larger feasibility region. In addition, even the stringent constant-envelope constraint is enforced on the waveforms, the SINR of the synthesized low-PAPR waveforms is very close to that of energy-constrained waveforms. Moreover, the SINR achieved by the DK-ADMM algorithm is slightly higher than that of the MM-ADMM algorithm.
Regarding the CPU time to reach convergence, as shown in Table \ref{Table:4}, the DK-ADMM algorithm is faster than the MM-ADMM algorithm. Interestingly, the results therein also imply that a larger PAPR results in a faster convergence.

\figurename~\ref{Fig:ESDs} presents the ESDs of the designed waveforms. The three stopbands are shaded in gray with red dash-dot lines. The blue lines indicate the ESDs of the initial waveforms, and the yellow lines denote the ESDs of the designed waveforms. From  \figurename~\ref{Fig:ESDs}, we can observe that all the transmit waveforms form deep nulls in the  stopbands and satisfy the spectral constraints. In other words, the designed waveform can precisely control the energy leaked on the stopbands, which enhance the coexistence between the radar system and other radio frequency systems. Moreover, we can observe that the ESDs of the waveforms synthesized by the DK-ADMM algorithm is smoother than by the MM-ADMM algorithm. Considering that the DK-ADMM algorithm achieves a larger SINR in a shorter time and the associated ESDs of the synthesized waveforms are smoother, we use the DK-ADMM algorithm to synthesize the multi-spectrally constrained waveforms in the sequel.

\begin{figure}[!htbp]
  \centering
  \subfigure[$\rho =1$]{\includegraphics[width = 0.24\textwidth]{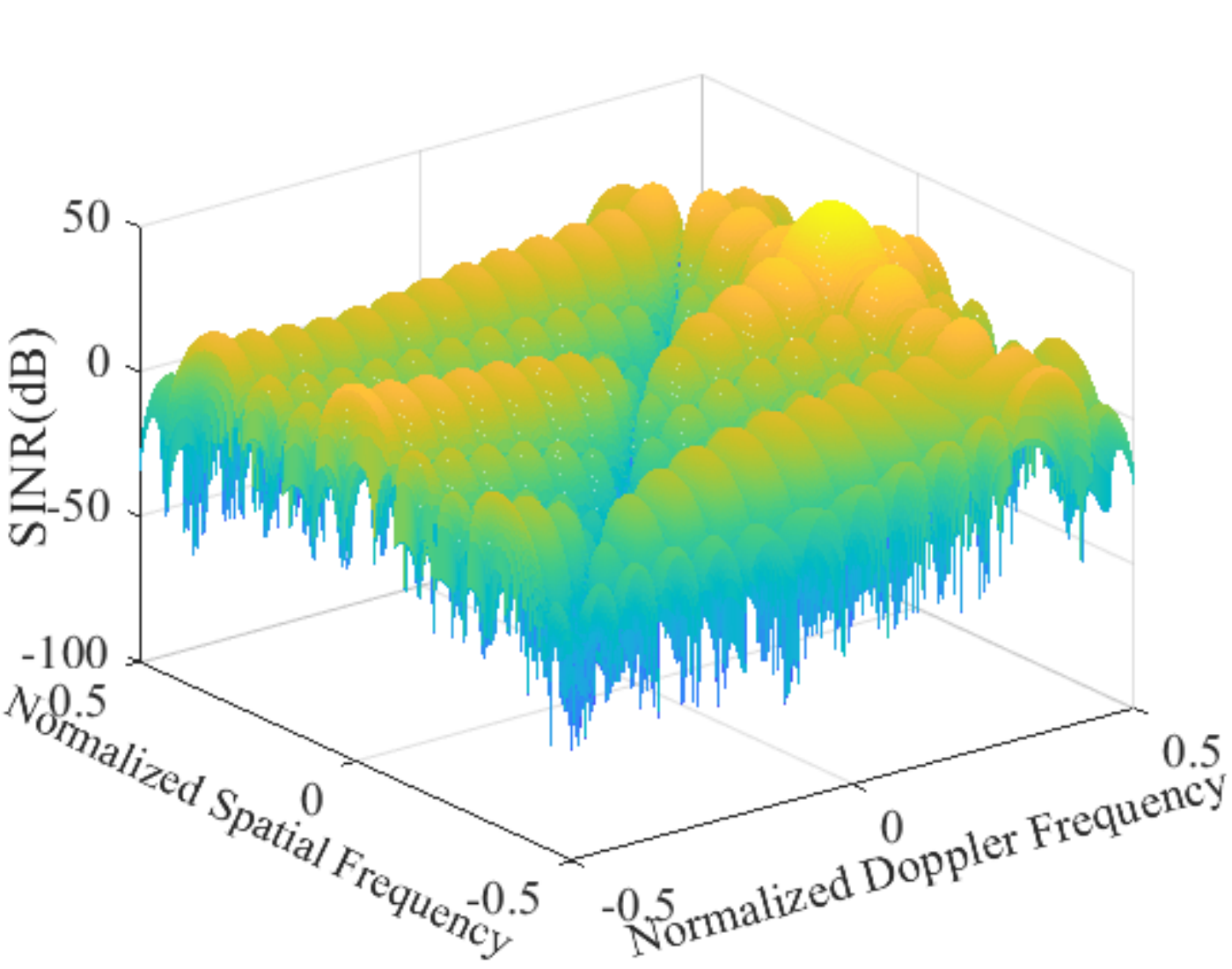}}
  \subfigure[$\rho =1$]{\includegraphics[width = 0.24\textwidth]{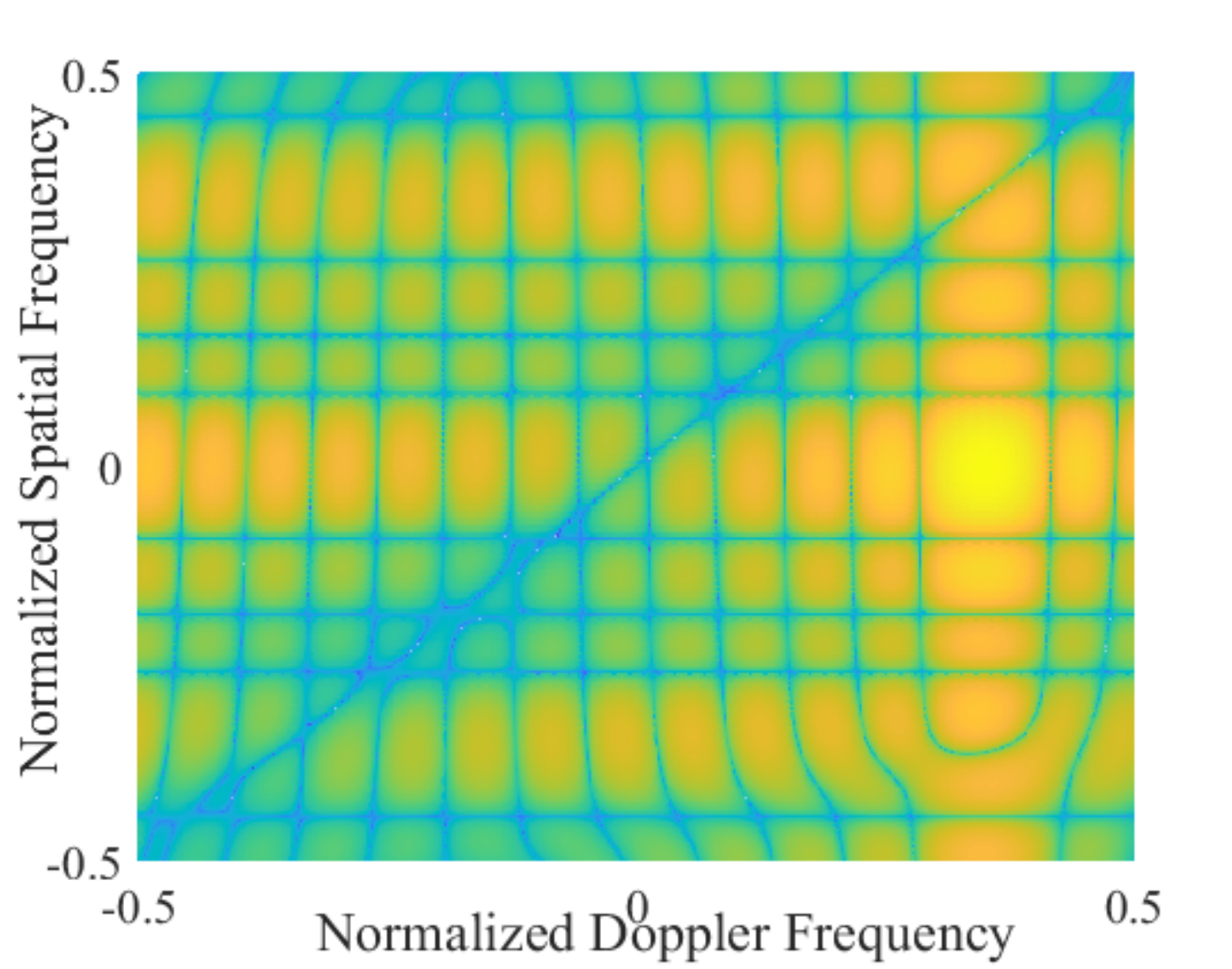}}
  \\ 
  \centering
  \subfigure[$\rho =L$]{\includegraphics[width = 0.24\textwidth]{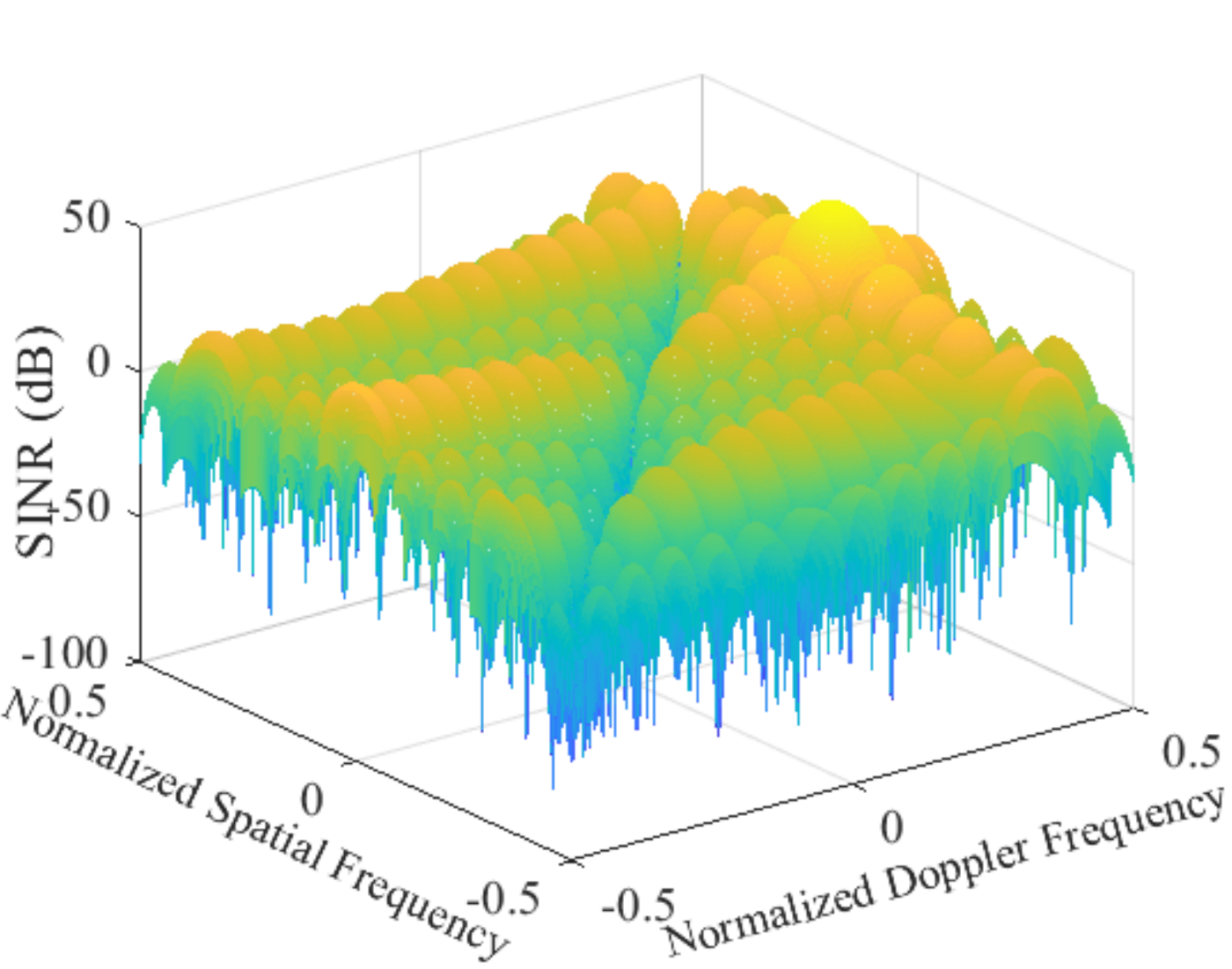}}
  \subfigure[$\rho =L$]{\includegraphics[width = 0.24\textwidth]{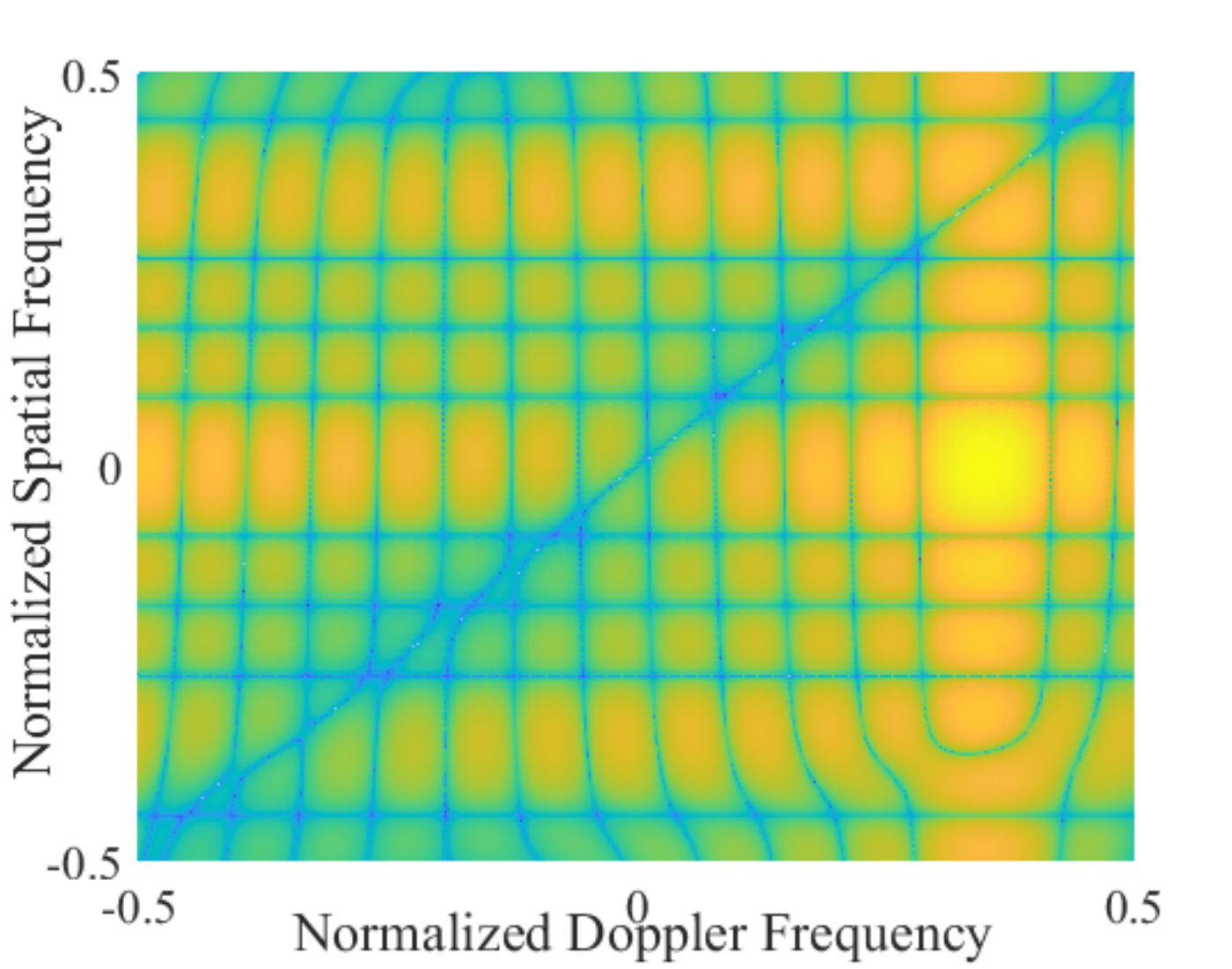}}

  \caption{STCA of the designed waveforms. $e_t=1$, $v_t=52.5$ m/s. $E_I^1=E_I^2=-35$ dB, $E_I^3=-30$ dB. (a) 3D STCA function of the constant-envelope waveforms. (b) Top view associated with (a). (c) 3D STCA function of the energy-constrained waveforms. (d) Top view associated with (c).}
  \label{Fig:STCA}
  \end{figure}

Next we analyze the space-time cross-ambiguity (STCA) function of the devised waveforms under different constraints, where the STCA function is defined as \cite{tang2016joint}
\begin{equation}
  P_{\bw,\bs}(\theta,f)=|\bw^\dagger \bV(\theta,f)\bs|^2,
\end{equation}
where $\bV(\theta,f)=\bd(f)\otimes \bI_L \otimes \bA(\theta)$, and $\bd(f)=[1,\cdots,e^{j2\pi (M-1)f}]^\top$.
\figurename ~\ref{Fig:STCA} shows the STCA function of the constant-envelope waveforms and the energy-constrained waveforms. We can observe the mainlobes of  all the STCA functions at zero spatial frequency (which corresponds to an azimuth of $0^\circ$) and a normalized Doppler frequency of $0.35$. Additionally, these functions form deep nulls along the clutter ridges. Therefore, the devised waveforms and filters can successfully suppress the clutter and improve the SINR performance. 

\begin{figure} [!htbp]
   \centering
   {\subfigure[]{\includegraphics[width = 0.4\textwidth]{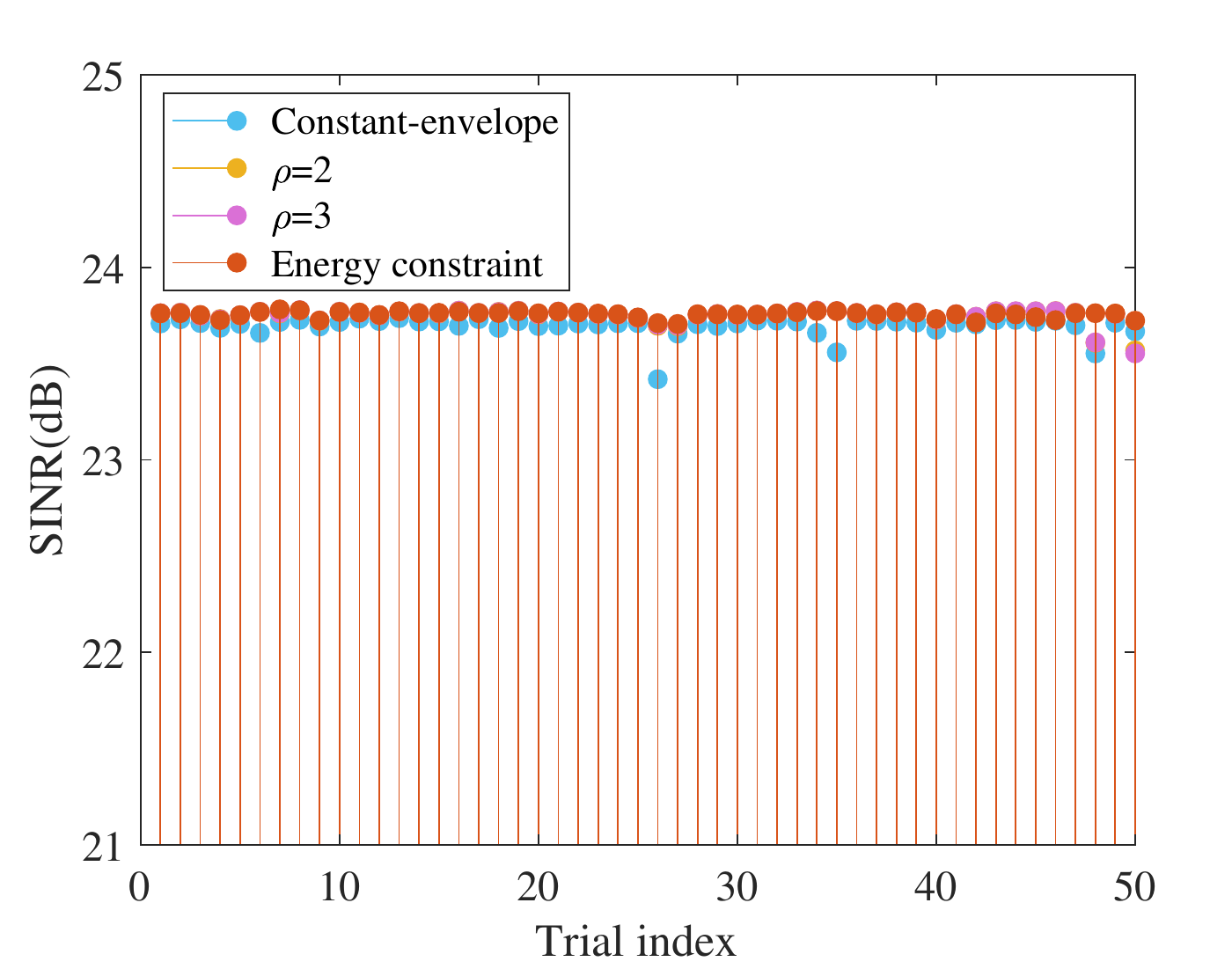}} \label{Fig:MonteCarlo.a}}
   {\subfigure[]{\includegraphics[width = 0.4\textwidth]{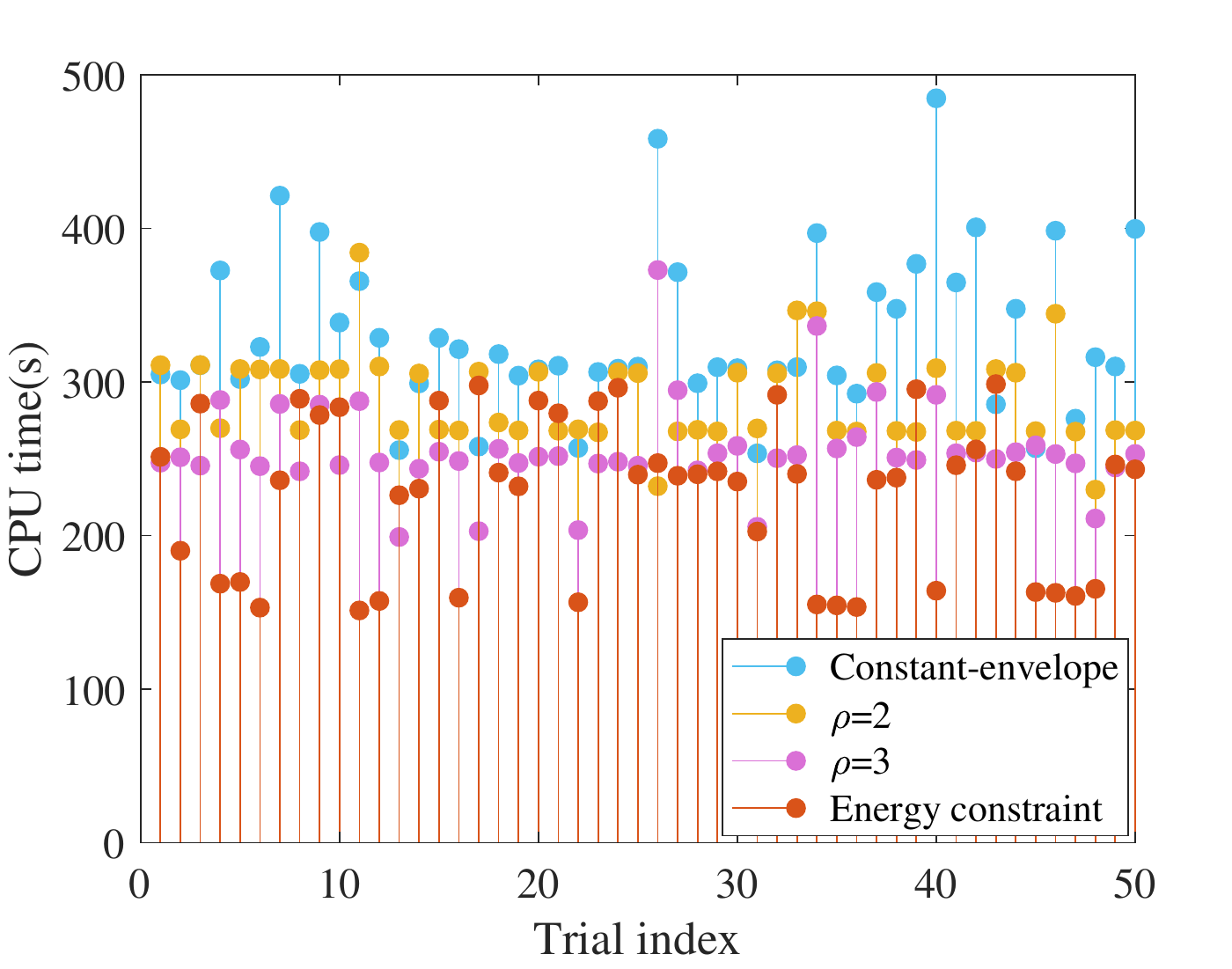}} \label{Fig:MonteCarlo.b}}
   \caption{ The impact of initial waveforms on SINR and CPU time. Random waveforms are used as the initial point. $e_t=1$, $v_t=52.5$ m/s. $E_I^1=E_I^2=-35$ dB, $E_I^3=-30$ dB. (a) SINR. (b) CPU time.}
   \label{Fig:MonteCarlo}
 \end{figure}

\begin{figure} [!htbp]
   \centering
   \subfigure[]{\includegraphics[width = 0.24\textwidth]{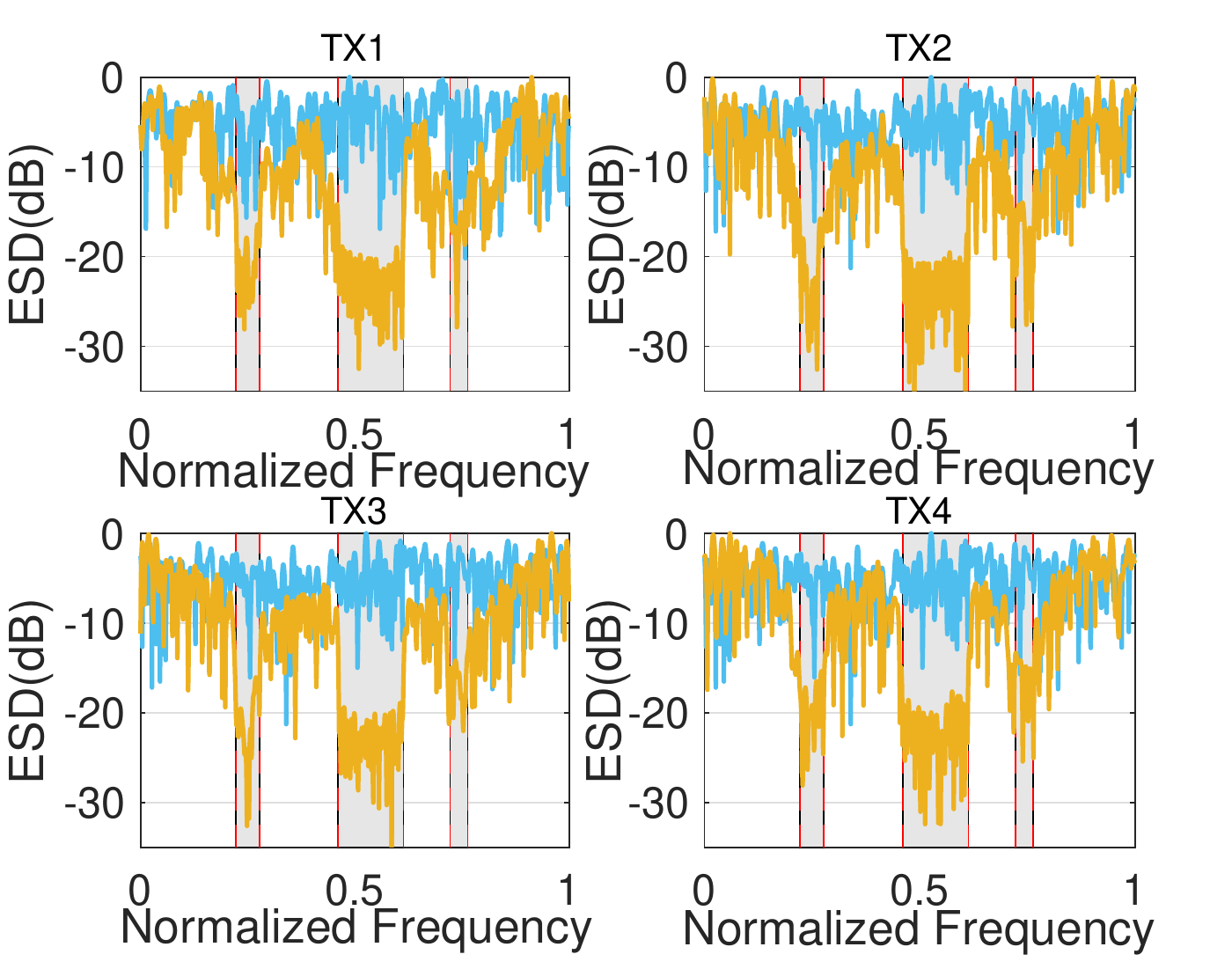}}
   \subfigure[]{\includegraphics[width = 0.24\textwidth]{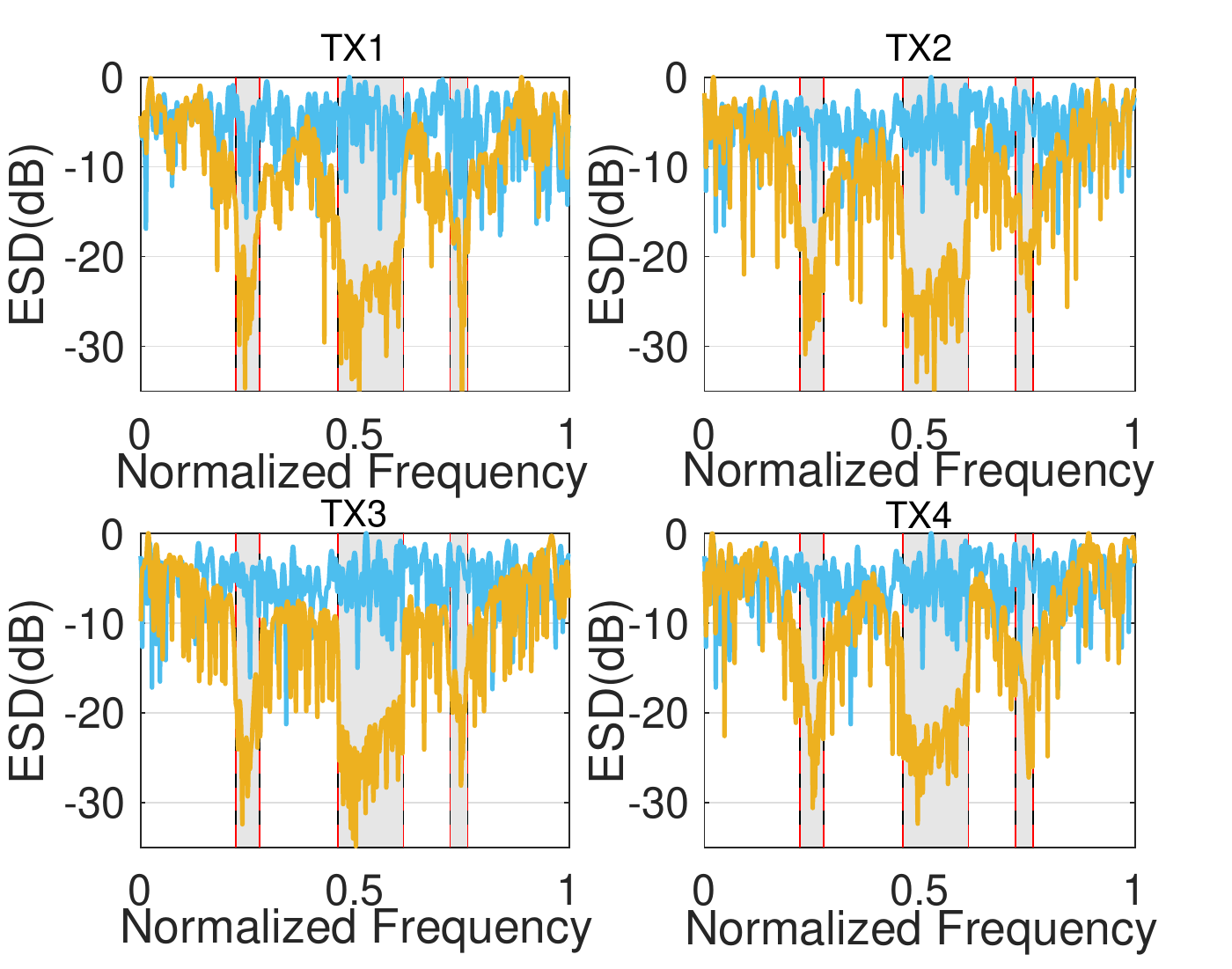}}
   \\ 
   \centering
   \subfigure[]{\includegraphics[width = 0.24\textwidth]{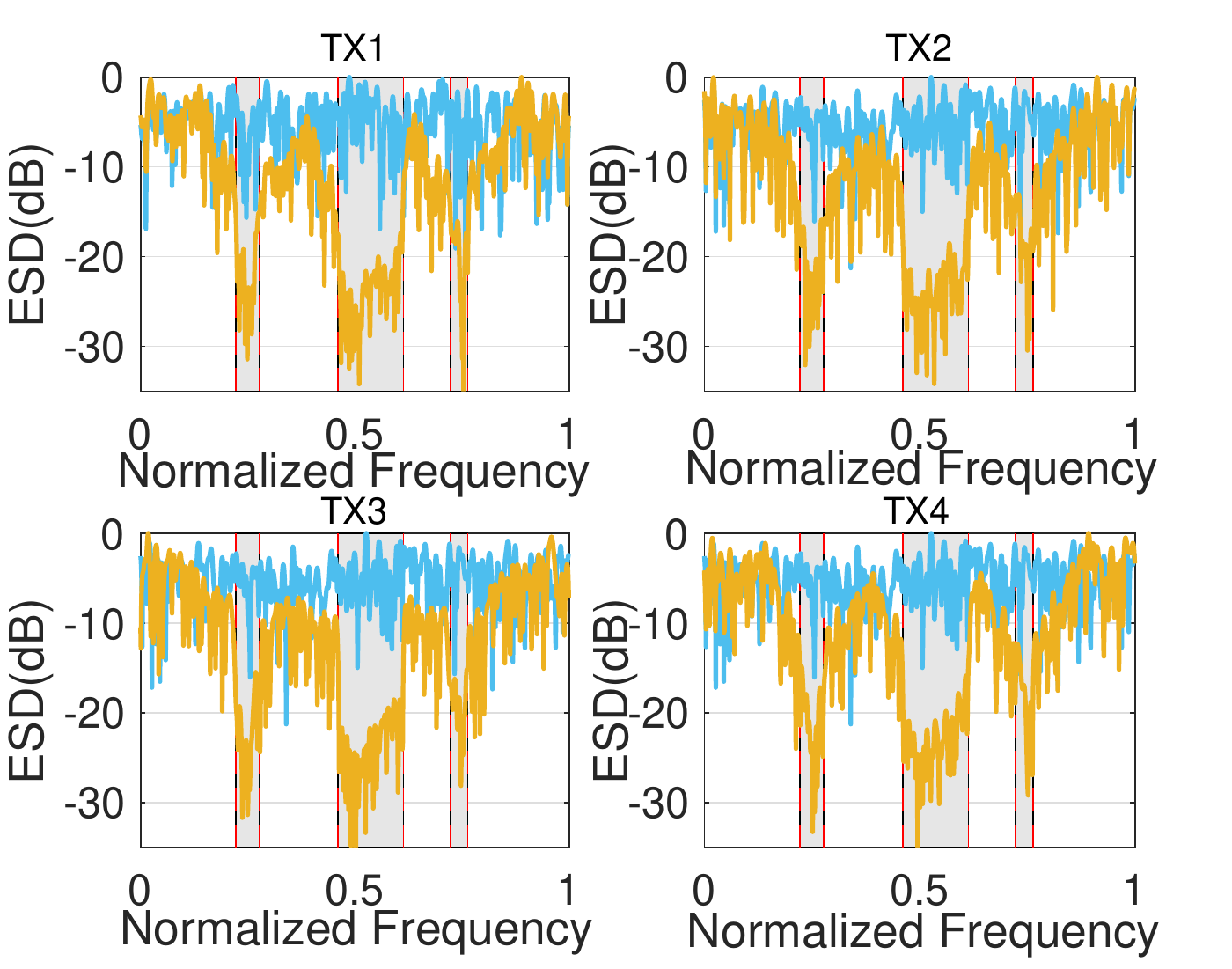}}
   \subfigure[]{\includegraphics[width = 0.24\textwidth]{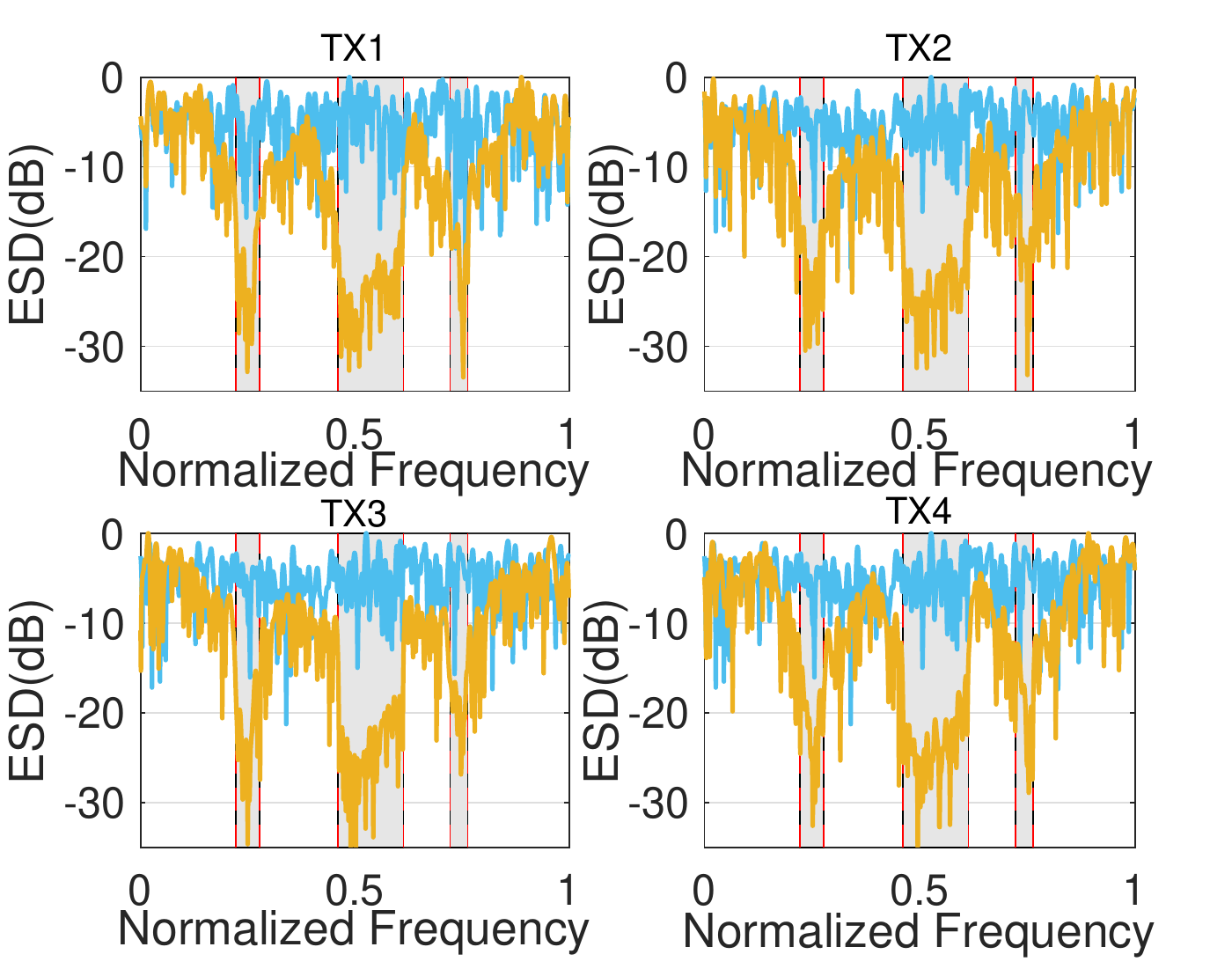}}
   \caption{ESDs of the designed waveforms. The blue and yellow lines represent the ESDs of the initial random waveforms and the optimized waveforms. $e_t=1$. $E_I^1=E_I^2=-35$ dB, $E_I^3=-30$ dB. (a) Constant-envelope waveforms. (b) $\rho=2$. (c) $\rho=3$. (d) Energy-constrained waveforms.}
   \label{Fig:random}
 \end{figure}

To assess the impact of initial points on the performance of the designed algorithms, various randomly generated waveforms are set to be the initial points, where the random waveforms are constant-envelope waveforms with modulus of $\sqrt{p_s}$ and phases following a zero-mean Gaussian distribution. The SINRs at convergence and the associated CPU time for different initial points are shown in \figurename ~\ref{Fig:MonteCarlo}, where 50 Monte Carlo trials are conducted. Table \ref{Table:5} and Table \ref{Table:6} show the maximum, the average, and the minimum value of the SINR at convergence and the CPU time needed to reach convergence for the different PAPR-constrained waveforms (i.e., $\rho=1,2,3,L$).
From \figurename ~\ref{Fig:MonteCarlo} and the results in Table \ref{Table:5} and \ref{Table:6}, we find that the SINR of the designed algorithm is insensitive to the initial points, but the convergence speed is affected by the initial points.
To show that the synthesized waveforms satisfy the spectral constraint, we randomly select a set of the results and plot the ESDs of the designed waveforms in \figurename ~\ref{Fig:random}. The results indicate that compared with the initial waveforms, the waveforms devised via the proposed algorithm achieve better spectral compatibility. Interestingly, the spectrum of the waveforms initialized by randomly generated waveforms is not as smooth as that in \figurename ~\ref{Fig:ESDs}.

\begin{table}[!htbp]
  \caption{{{SINR at convergence}}}
  \centering
  \begin{tabular}{c c c c}
   \hline
    SINR (dB) & Maximum & Average & Minimum \\
   \hline
   Constant envelope& 23.736 & 23.696 &  23.419 \\
   $\rho=2$& 23.777 & 23.750 & 23.569 \\
   $\rho=3$& 23.778 & 23.751 & 23.553 \\
   Energy constraint& 23.782 & 23.755 & 23.706 \\
   \hline
   \end{tabular}
   \label{Table:5}
\end{table}

\begin{table}[!htbp]
    \caption{{{CPU time needed to reach convergence}}}
    \centering
    \begin{tabular}{c c c c}
     \hline
      CPU time (s) & Maximum & Average & Minimum \\
     \hline
     Constant envelope& 484.717 & 329.859 & 253.514 \\
     $\rho=2$& 384.215 & 288.868 & 229.881 \\
     $\rho=3$& 372.878 & 255.552 & 199.061 \\
     Energy constraint& 298.581 & 227.069 & 151.206 \\
     \hline
     \end{tabular}
     \label{Table:6}
\end{table}

\begin{figure}[!htbp]
    \centering
    \subfigure[]{\includegraphics[width=0.35\textwidth]{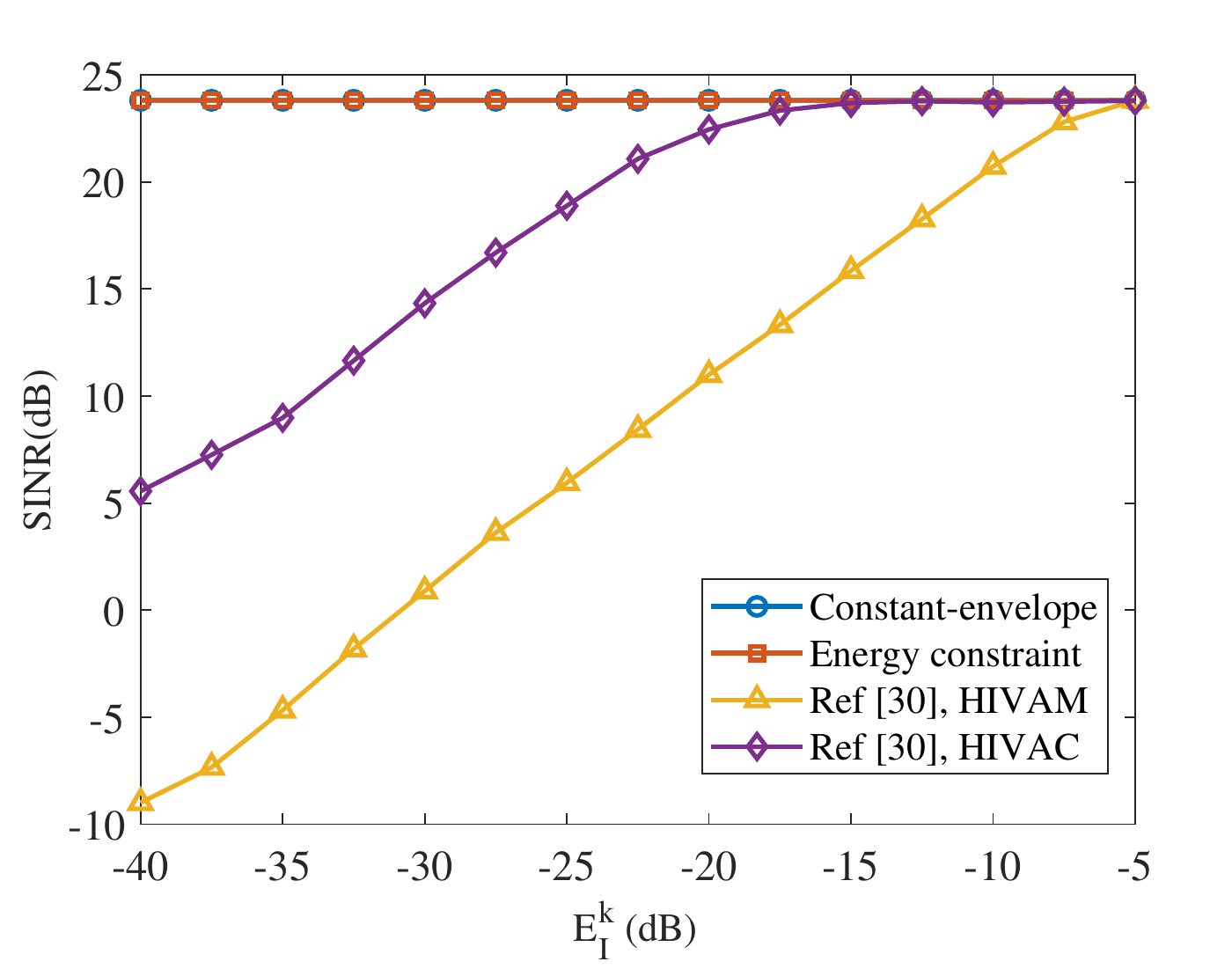}}
    \subfigure[]{\includegraphics[width=0.35\textwidth]{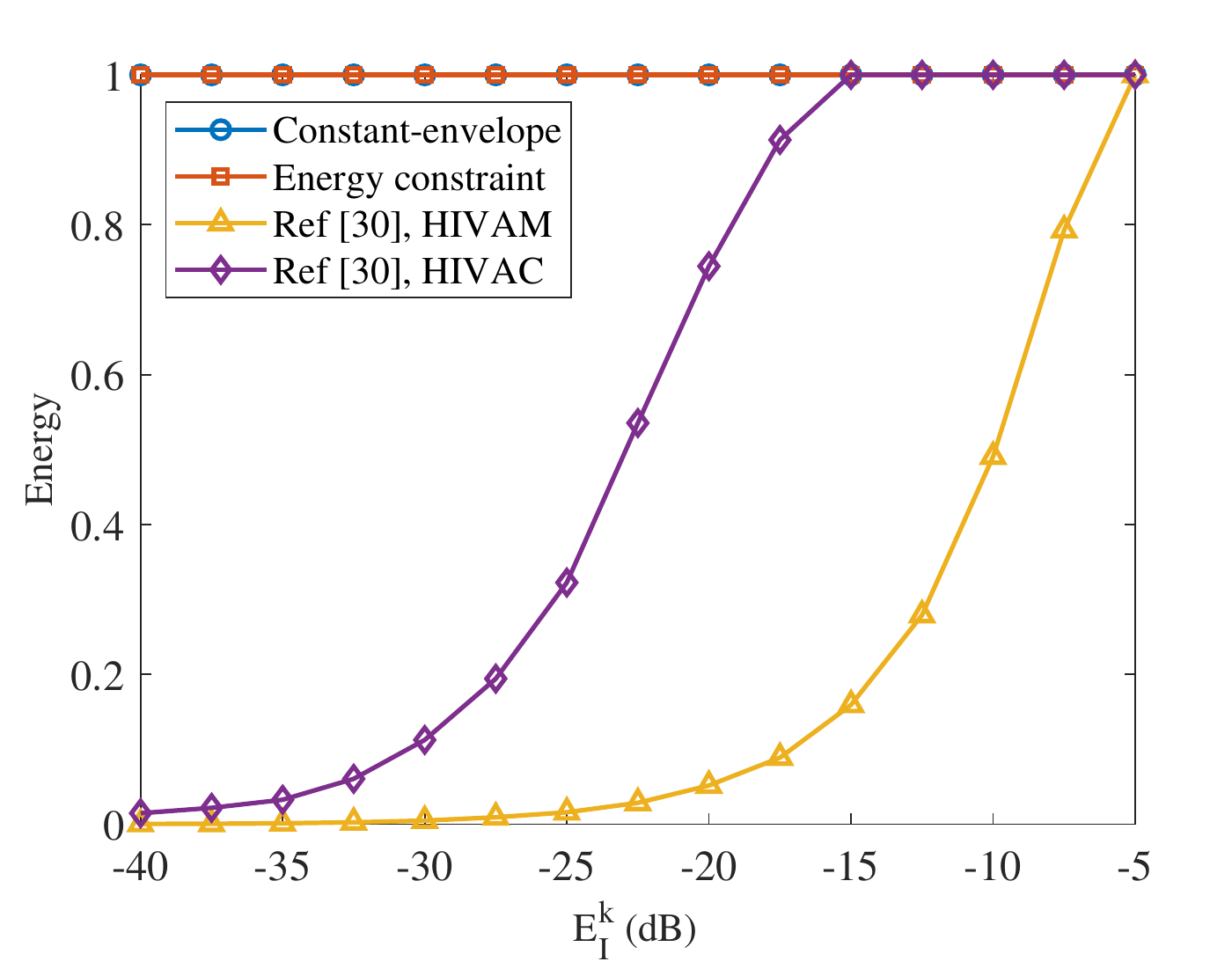}}
    \caption{(a) The impact of $E_I^k$ on SINR. (b) The energy of the waveforms at convergence. $e_t=1$, $v_t=52.5$m/s. $E_I^1=E_I^2=E_I^3$.}
    \label{Fig:EIK}
\end{figure}

In the following, we analyze the impact of spectral notch depths on the achieved SINR. \figurename ~\ref{Fig:EIK}(a) shows the achieved SINR with respect to different spectral notch depths and compares with that of the waveforms devised via the algorithm in \cite{Yang2022Multispectrally} \footnote{It should be noted that the algorithm in \cite{Yang2022Multispectrally} focuses on designing constant-envelope waveforms for a SISO radar system. Herein, we extend this algorithm to deal with the MIMO case. However, it is difficult for the algorithm in \cite{Yang2022Multispectrally} to secure a feasible initial point to satisfy both the equality constraint $\bs^\dagger\bs =e_t$ and the multi-spectral constraint. Therefore, when using the algorithm  in \cite{Yang2022Multispectrally} to design the constant-envelope waveforms, we replace this  equality constraint  with the inequality constraint $\bs^\dagger\bs \leq e_t$.},  which is initialized by a heuristic initialization via alternating optimization with MM (HIVAM) or a  heuristic initialization via alternating optimization with CD (HIVAC) method.
For simplicity we assume that $E_I^1=E_I^2=E_I^3 = E_I$. It can be seen that as the notch depth goes deeper, the waveforms devised via the proposed algorithm attain higher SINR than those synthesized by the algorithm in \cite{Yang2022Multispectrally}. This is because that the algorithm in \cite{Yang2022Multispectrally} needs to scale the energy of the waveforms to satisfy the multi-spectral constraint. To see this, \figurename ~\ref{Fig:EIK}(b) draws the energy of the waveforms. It can be observed that as the notch depth goes deeper, the energy of the waveforms devised via the algorithm in \cite{Yang2022Multispectrally} drops to a low level (to satisfy the stringent multi-spectral constraint), while the energy of the waveforms devised via the proposed algorithm always reaches the highest possible level. Since the SINR performance improves with the waveform energy, the performance of our waveforms is superior to that of the waveforms devised via the algorithm in \cite{Yang2022Multispectrally}.

\begin{figure} [!htbp]
  \centering
  {\subfigure[]{\includegraphics[width = 0.35\textwidth]{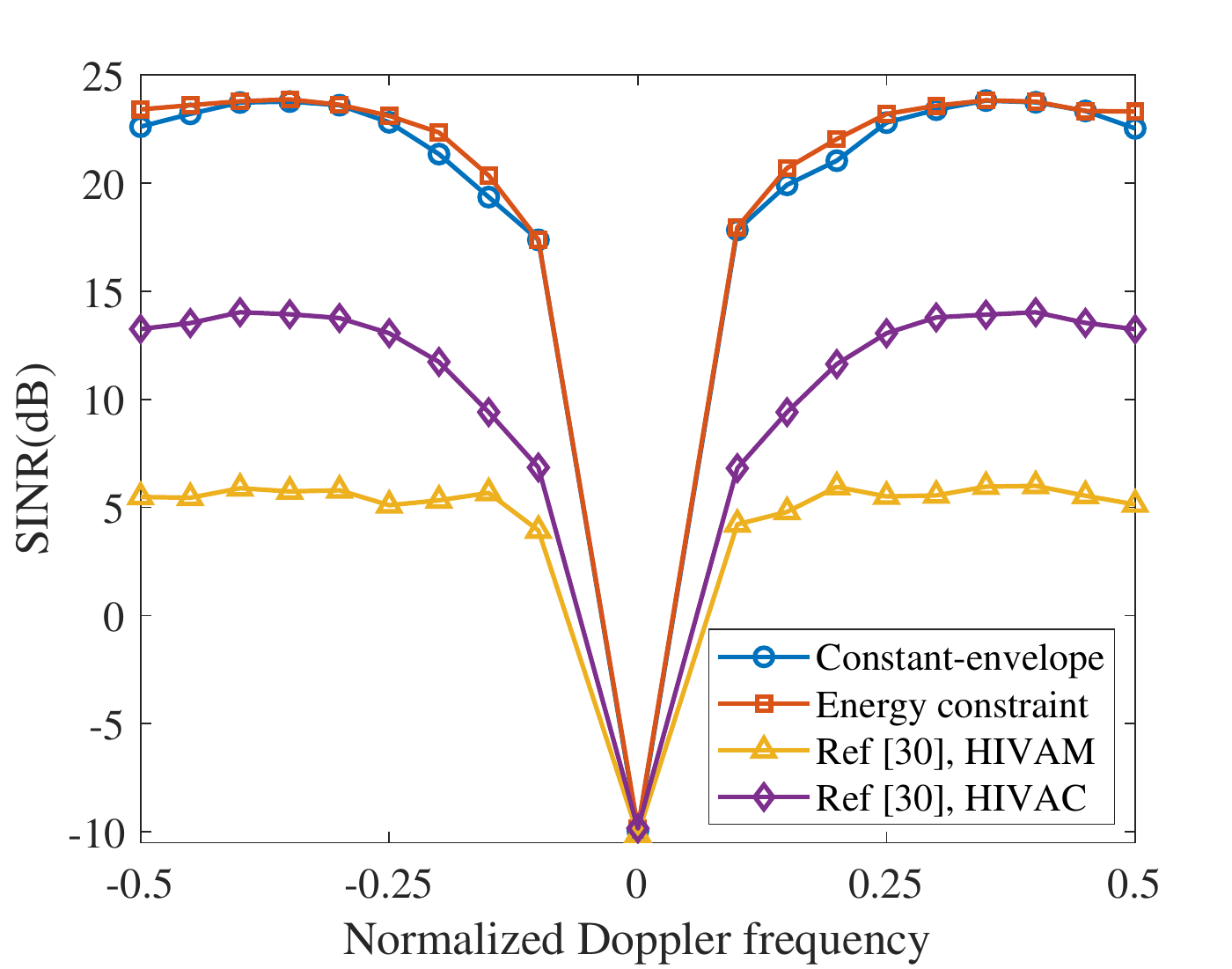}}}
  {\subfigure[]{\includegraphics[width = 0.35\textwidth]{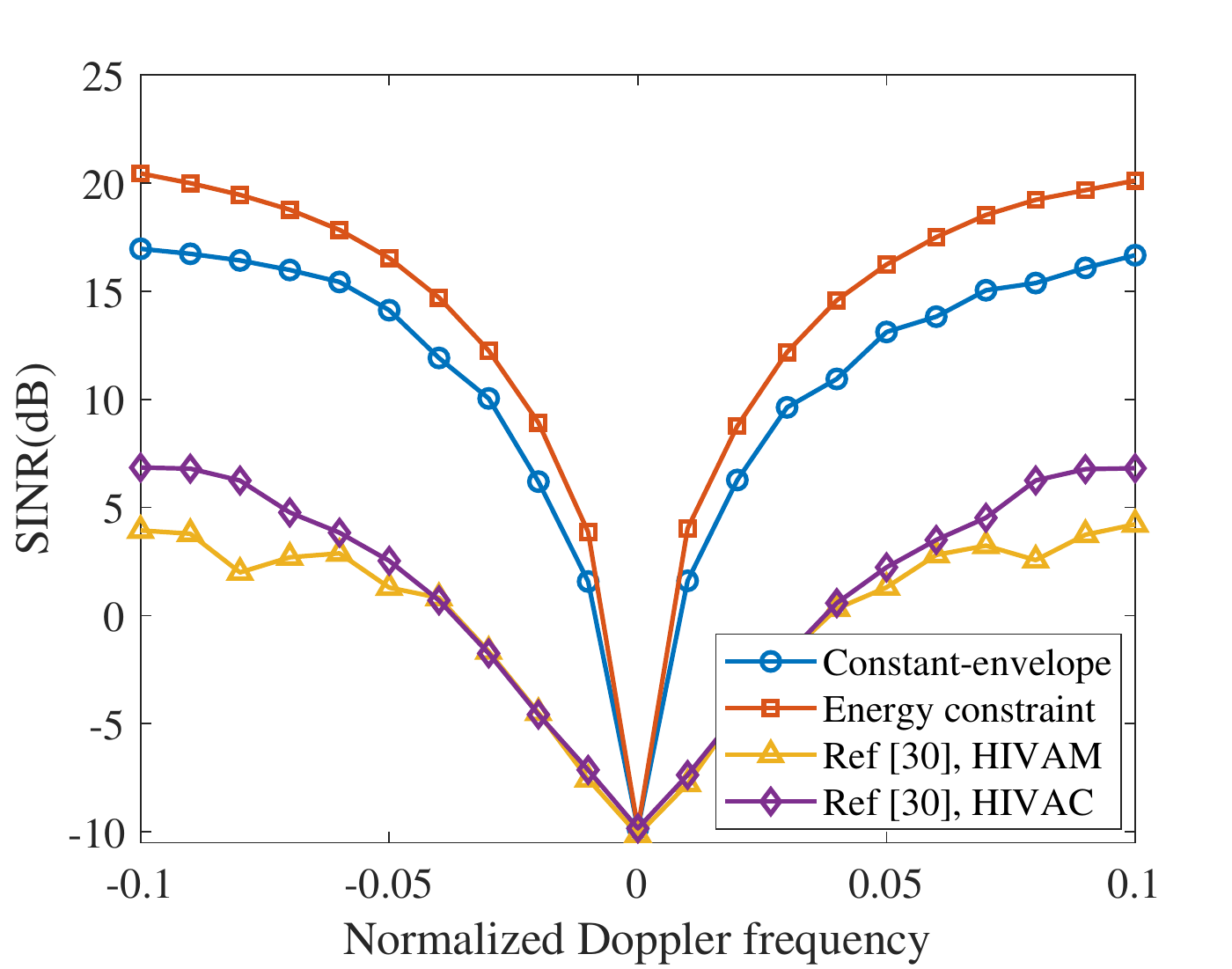}}}
  \caption{(a) SINR versus normalized target Doppler frequency. $f_t\in [-0.5,0.5]$. (b) SINR at low Doppler frequencies. $f_t\in [-0.1,0.1]$. $E_I^1=E_I^2=-35$ dB, $E_I^3=-30$ dB.}
  \label{Fig:Doppler}
\end{figure}

\figurename~\ref{Fig:Doppler} compares the SINR of the waveforms devised via the proposed algorithm versus the normalized target Doppler frequencies with the algorithm in \cite{Yang2022Multispectrally}, where the performance of the energy-constrained waveforms is also included as a benchmark.
From \figurename ~\ref{Fig:Doppler}, we can see that the waveforms devised via the proposed algorithm achieve better detection performance than those devised via the algorithm in \cite{Yang2022Multispectrally}, especially at the low Doppler frequency area. 

\begin{figure} [!htbp]
  \centering
  {\subfigure[]{\includegraphics[width = 0.35\textwidth]{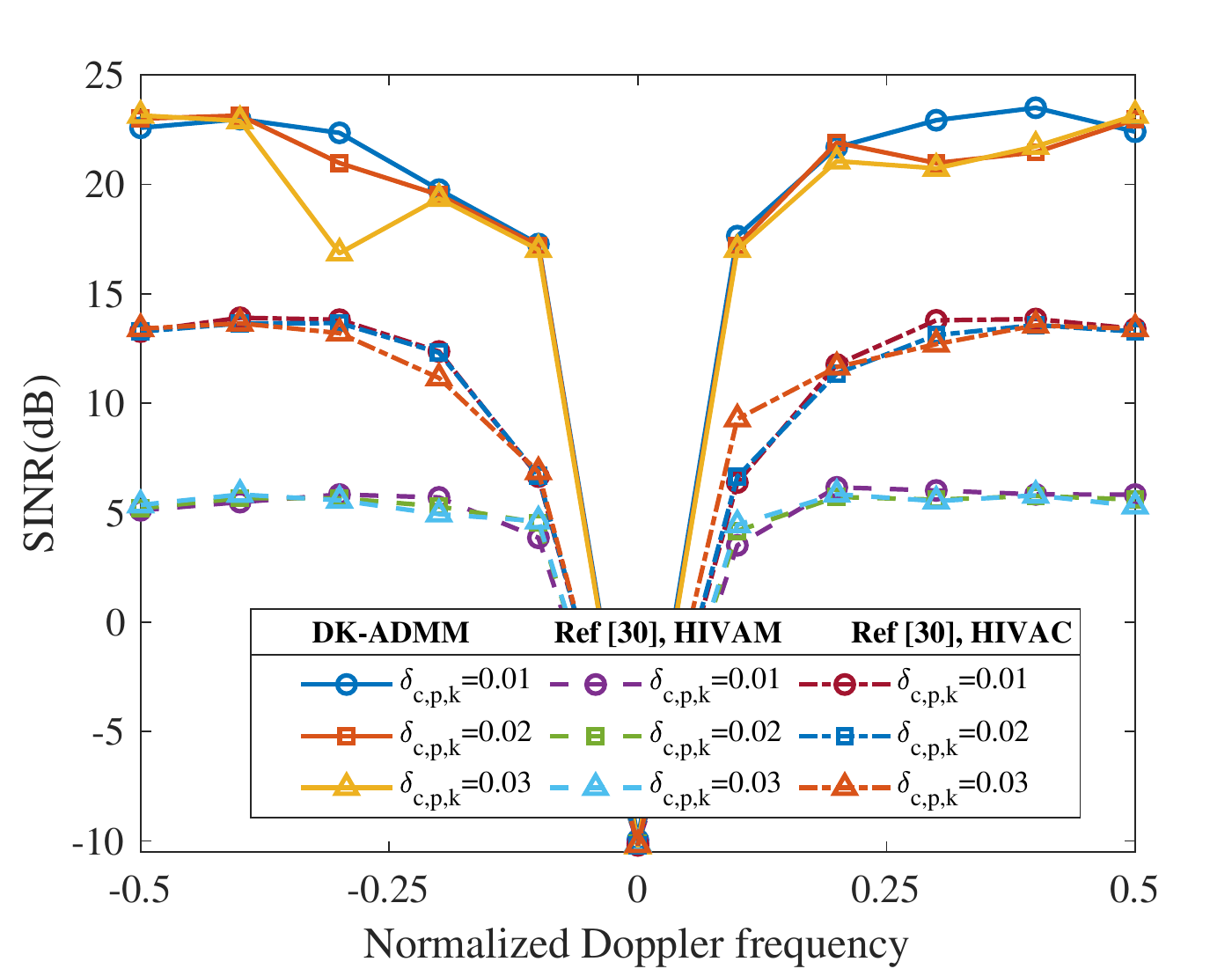}}}
  {\subfigure[]{\includegraphics[width = 0.35\textwidth]{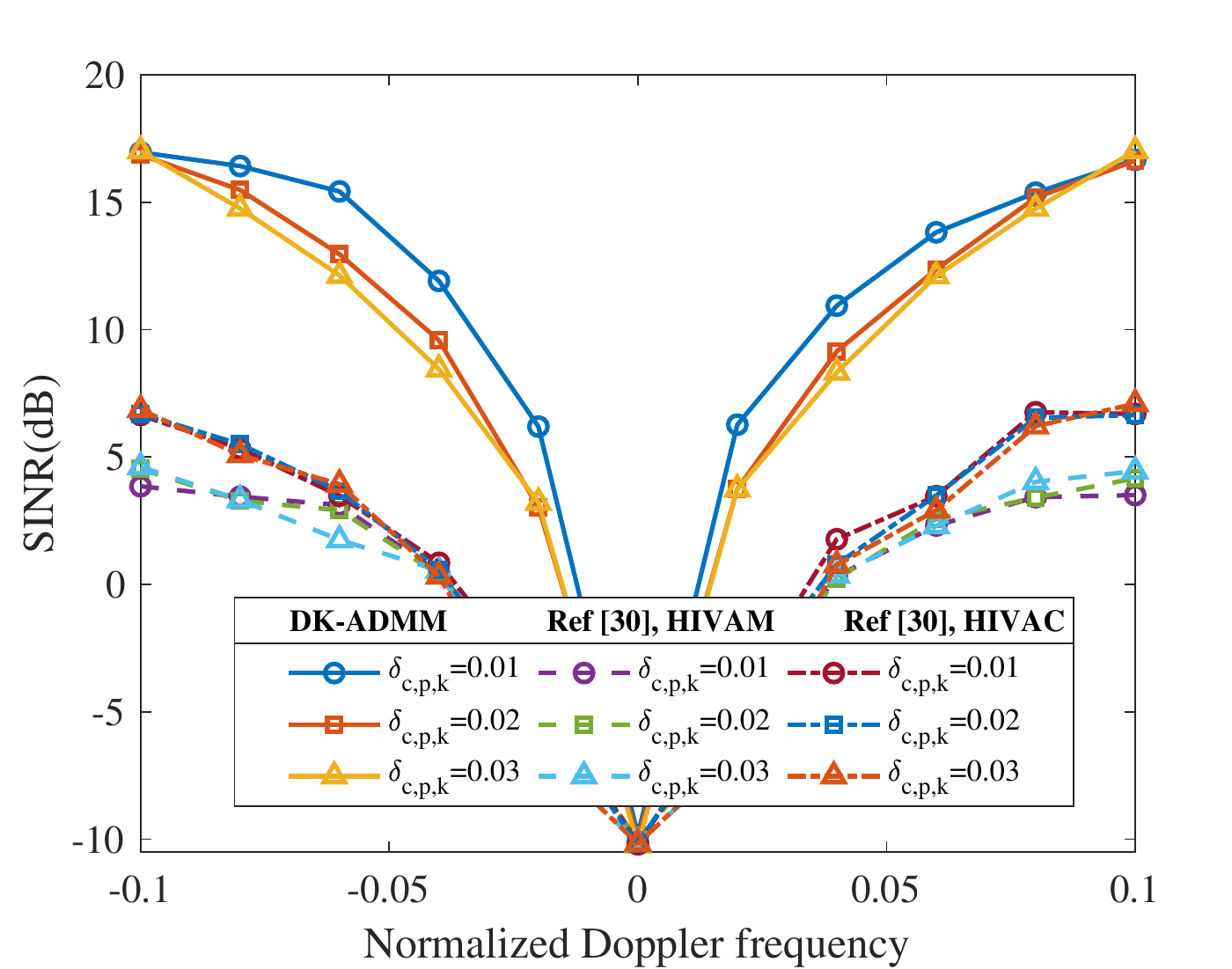}}}
  \caption{(a) SINR versus normalized target Doppler frequency. (b) SINR at low Doppler frequencies. $f_t\in [-0.5,0.5]$, $E_I^1=E_I^2=-35$ dB, $E_I^3=-30$ dB. }
  \label{Fig:Doppler_uncertainty2}
\end{figure}

Next, we assess the robustness of the proposed algorithm with respect to the Doppler uncertainty of the clutter patches.
\figurename ~\ref{Fig:Doppler_uncertainty2} shows the SINR of the constant-envelope  waveform versus the normalized Doppler frequency under different clutter uncertainty. Note that the Doppler uncertainty degrades the  target detection performance, especially in the low Doppler frequency region (about $2\sim3$ dB loss in this area). However, the proposed algorithm still achieves better radar detection performance than the competing algorithms.

\begin{figure} [!htbp]
  \centering
  \includegraphics[width = 0.35\textwidth]{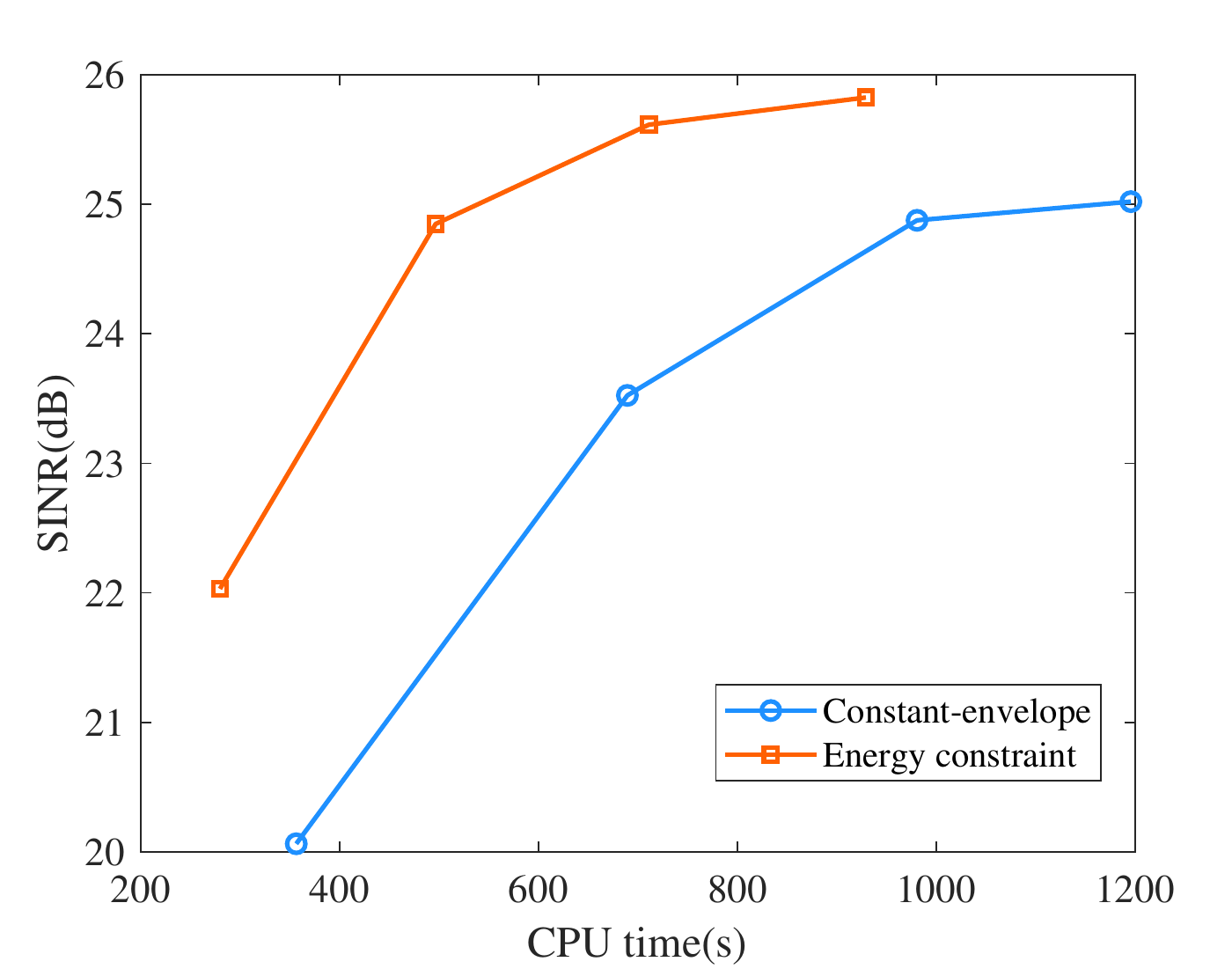}
  \caption{Convergence of SINR with respect to CPU time. $d_t=\lambda/2$, $d_r=2\lambda$. $M=24$. $\phi=10^\circ$. $E_I^1=E_I^2=E_I^3=-35$ dB. $\Theta_1 = [-60^\circ, -25^\circ]$, $\Theta_2 = [20^\circ, 60^\circ]$, $\Theta_1 = [25^\circ, 70^\circ]$}
  \label{Fig:space-frequency_convergence}
\end{figure}

\begin{figure} [!htbp]
  \centering
  \subfigure[]{\includegraphics[width = 0.24\textwidth]{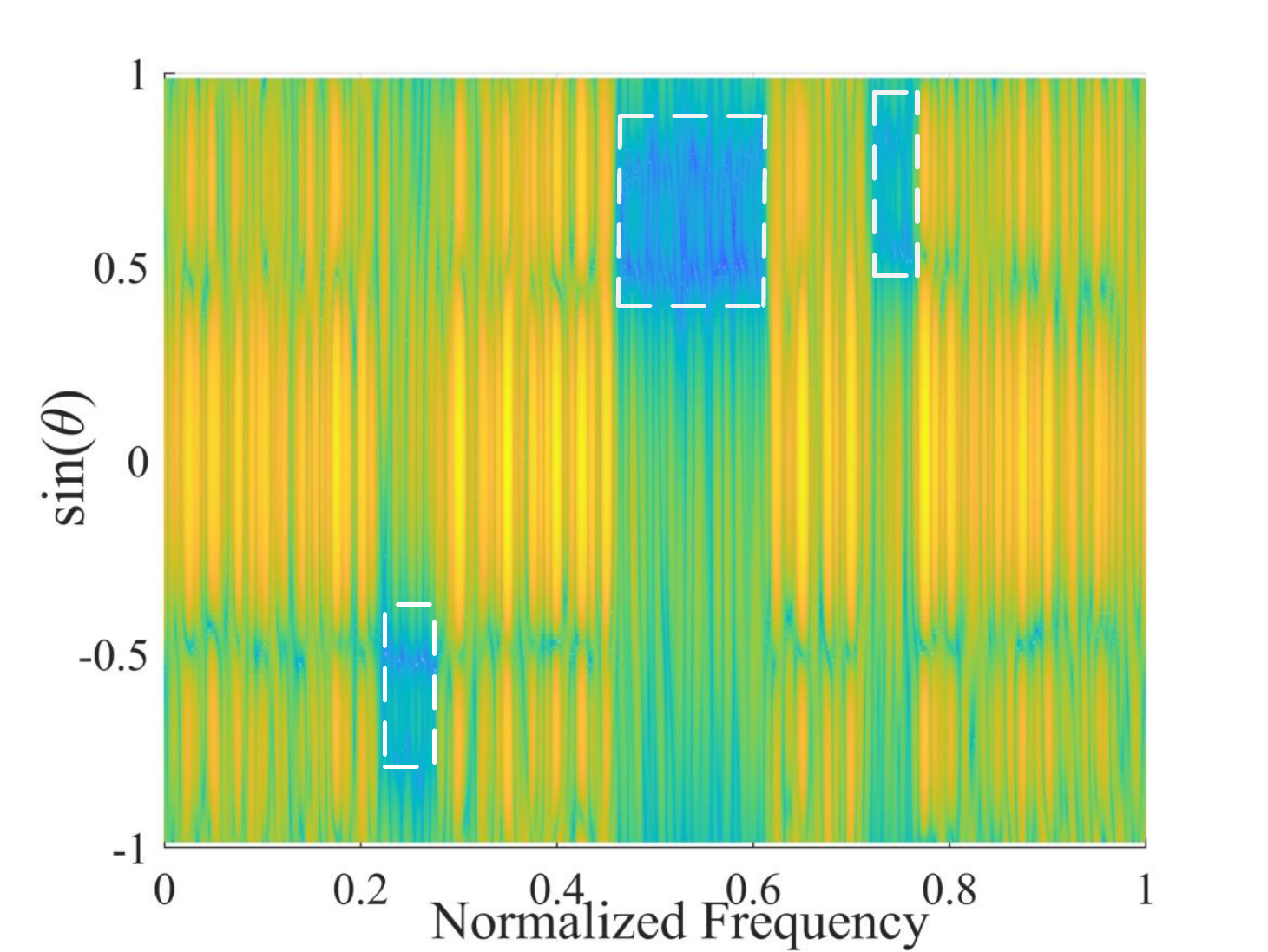}}
  \subfigure[]{\includegraphics[width = 0.24\textwidth]{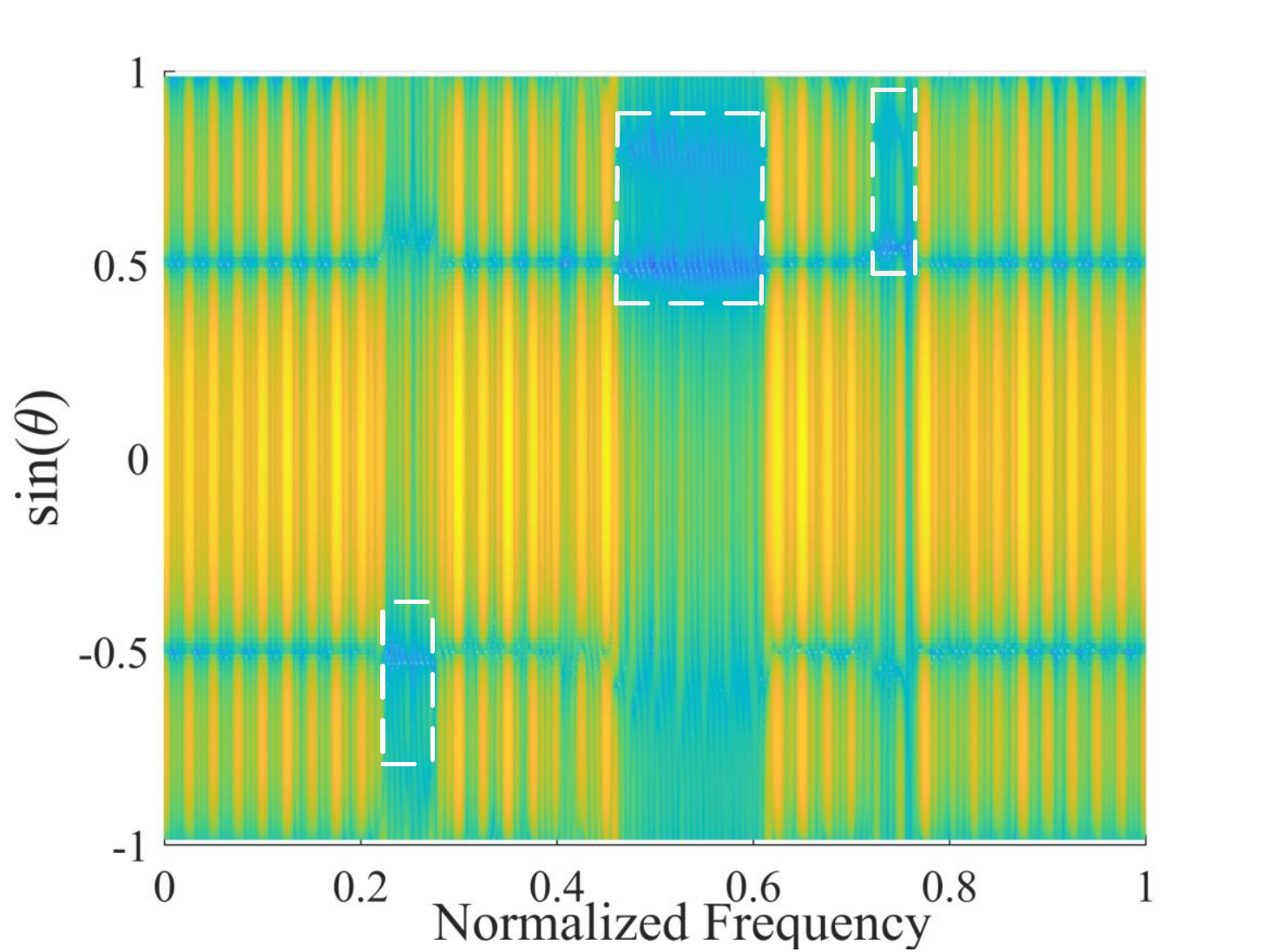}}
  \caption{The spectral distributions on the spatial-frequency domain. $d_t=\lambda/2$, $d_r=2\lambda$. $M=24$. $\phi=10^\circ$. $E_I^1=E_I^2=E_I^3=-35$ dB. $\Theta_1 = [-60^\circ, -25^\circ]$, $\Theta_2 = [20^\circ, 60^\circ]$, $\Theta_1 = [25^\circ, 70^\circ]$. (a) Constant-envelope waveforms (b) Energy-constrained waveforms. }
  \label{Fig:space-frequency}
\end{figure}

Finally, we extend the proposed ADMM algorithm to deal with the multiple space-frequency constraints.
\figurename ~\ref{Fig:space-frequency_convergence} analyzes the SINR of the proposed algorithm versus the CPU time, where the spatial regions associated with the three radiators are $\Theta_1 = [-60^\circ, -25^\circ]$, $\Theta_2 = [20^\circ, 60^\circ]$, and $\Theta_1 = [25^\circ, 70^\circ]$, respectively, $E_I^1=E_I^2=E_I^3=-35$ dB. The inter-element spacing is set to be $d_t=\lambda/2$ and $d_r=2\lambda$. The elevation of the target of interest is set to be $\phi=10^\circ$. The MIMO radar transmits $M=24$ pulses in a CPI. The results show the monotonically increasing SINR of the waveforms synthesized by the proposed algorithm.
\figurename~\ref{Fig:space-frequency} shows the spectral distribution of the synthesized waveforms over the spatial-frequency domain. We can see that the synthesized waveforms can precisely control the energy leaked on the spatial-frequency domains corresponding to the radiators, further improving the spectral coexistence of the MIMO radar system and the nearby radiators.

\section{Conclusions}
We derived efficient algorithms to design low-PAPR waveforms for airborne MIMO radar in spectrally crowded environments. The purpose was to maximize the output SINR by jointly optimizing the transmit waveforms and receive filters. To tackle the multi-spectrally constrained waveform optimization problem, we developed two iterative algorithms. which were based on cyclic optimization, Dinkelbach's transform, MM, and ADMM. Results showed that the waveforms devised via the proposed algorithm not only  improved the detection performance of airborne MIMO radar, but also attained better spectral compatibility.

Possible future work includes the design of filter banks to account for unknown target Doppler (see, e.g., \cite{Aubry2015Optimizing} for a discussion on this topic), the investigation of the correlation properties of the designed waveforms, and the performance analysis of the waveforms on hardware.  It's also crucial to develop computationally efficient algorithms to design the waveforms in real time. Finally, the theoretical analysis for the convergence of the proposed ADMM algorithm will be left as a future topic.

\ifCLASSOPTIONcaptionsoff
  \newpage
\fi
\bibliographystyle{IEEEtran}
\bibliography{IEEEabrv,reference}
\newpage

\section{Biography Section}

\begin{IEEEbiography}[{\includegraphics[width=1in,height=1.25in,clip,keepaspectratio]{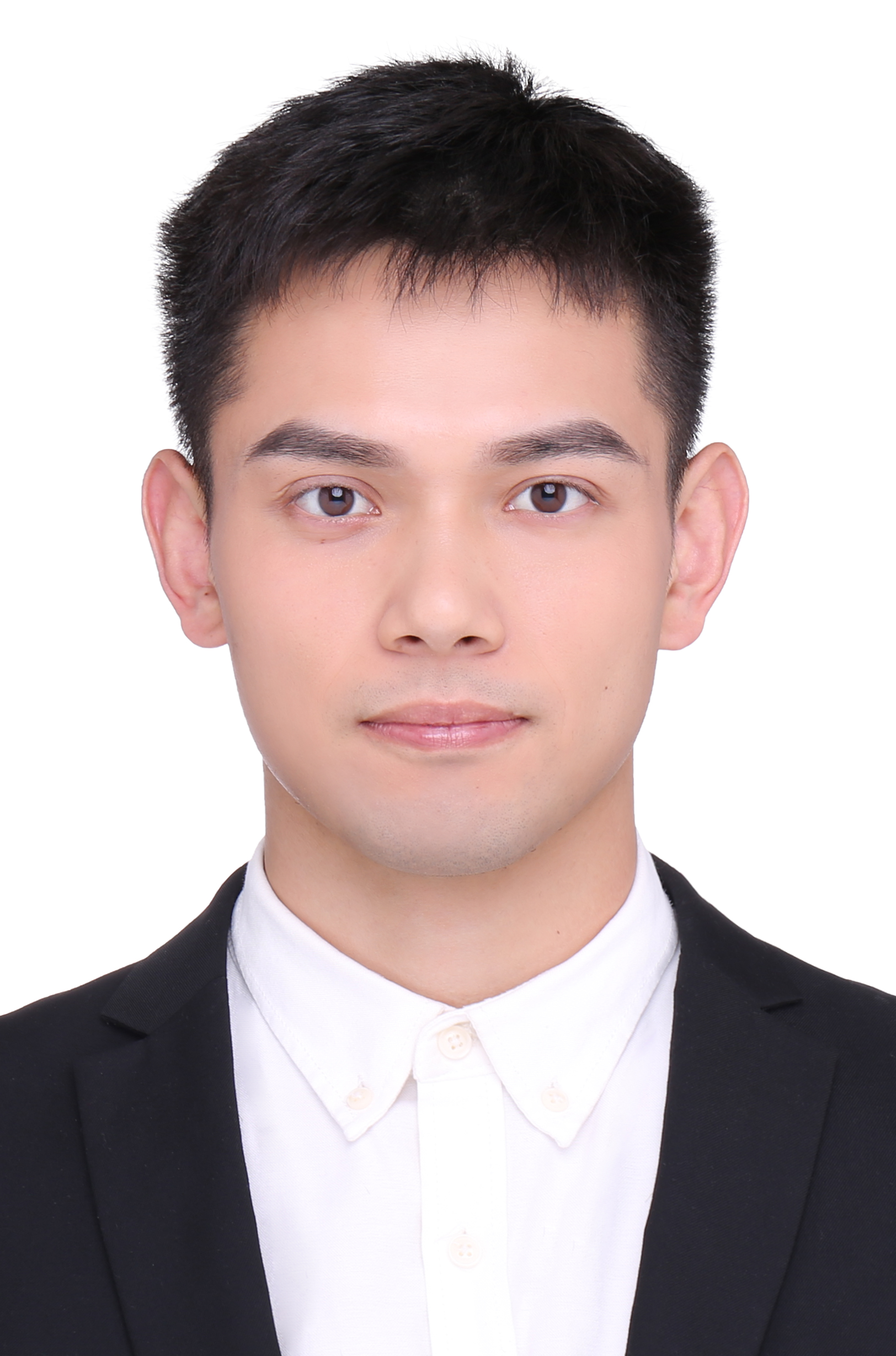}}]{Da Li}{\space}was born in Luoyang, Henan Province, China, in 1995. He received the B.S. degree from Lanzhou University, Lanzhou, China, in 2018, and the M.S. degree in 2020 from National University of Defense Technology, Hefei, China, where he is currently working toward the Ph.D. degree in information and communication engineering at the Department of College of Electronic Engineering. His research interests mainly include signal processing and radar waveform design.
\end{IEEEbiography}

\begin{IEEEbiography}[{\includegraphics[width=1in,height=1.25in,clip,keepaspectratio]{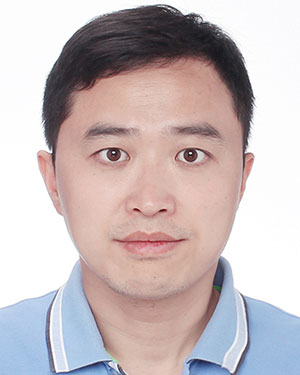}}]{Bo Tang} was born in Linchuan, Jiangxi Province, China, in 1985. He received the B.S. and Ph.D. degrees in electrical engineering from Tsinghua University, Beijing, China, in 2006 and 2011, respectively. From July 2011 to June 2017, he was with Electronic Engineering Institute, as a Lecturer. Since July 2017, he has been with the College of Electronic Engineering, National University of Defense Technology, Hefei, China, where he is currently a Professor. His research interests mainly include adaptive radar signal processing and radar waveform design. He was selected as the “Young Elite Scientists Sponsorship Program” by China Association for Science and Technology and sponsored by the Anhui Provincial Natural Science Foundation for Distinguished Young Scholars. He is currently an Associate Editor for the IEEE TRANSACTIONS ON SIGNAL PROCESSING.
\end{IEEEbiography}

\begin{IEEEbiography}[{\includegraphics[width=1in,height=1.25in,clip,keepaspectratio]{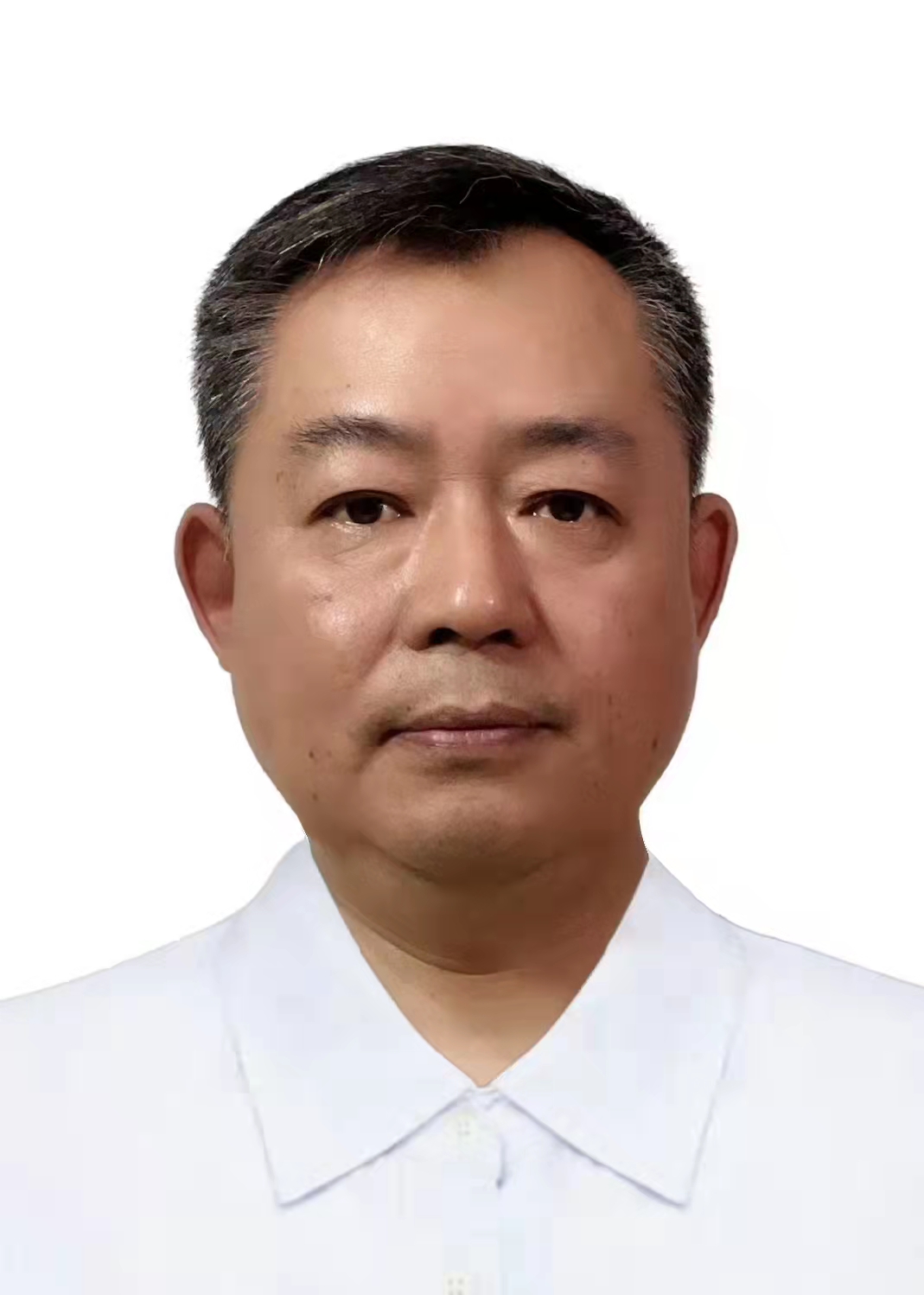}}]{Lei Xue}{\space}was born in Shouxian, Anhui, China, in 1963. He is currently a Professor of Information and Communication Engineering from the College of Electronic Engineering, National University of Defense Technology.
\end{IEEEbiography}

\end{document}